\documentclass[12pt,preprint]{aastex}
\usepackage{natbib}

\shorttitle{SMA Survey of Protoplanetary Disks in Orion}
\shortauthors{Mann \& Williams}

\received{2010}
\begin{document}

\title{A Submillimeter Array Survey of Protoplanetary Disks in the Orion Nebula Cluster}

\author{Rita K. Mann\altaffilmark{1,2} \and Jonathan P. Williams\altaffilmark{2}}
\altaffiltext{1}{National Research Council Canada, Herzberg Institute of Astrophysics,
5071 West Saanich Road, Victoria, British Columbia, Canada V9E 2E7} 
\altaffiltext{2}{Institute for Astronomy, University of Hawaii, 
2680 Woodlawn Drive, Honolulu, HI 96822}

\email{rita.mann@nrc-cnrc.gc.ca,jpw@ifa.hawaii.edu}

\begin{abstract}
We present the full results of our 3-year long Submillimeter Array
%\footnote{The Submillimeter Array is a joint
%project between the Submillimeter Astrophysical Observatory and the Academica
%Sinica Institute of Astronomy and Astrophysics and is funded by the Smithsonian
%Institution and the Academica Sinica.}
survey of protoplanetary disks in the Orion Nebula Cluster.
We imaged 23 fields at $880\,\mu$m and 2 fields at $1330\,\mu$m,
covering an area of $\sim$\,6.5 arcmin$^2$ and containing 67 disks.
We detected 42 disks with fluxes between 6-135 mJy and at
rms noise levels between 0.6 to 5.3 mJy beam$^{-1}$.
Thermal dust emission above any free-free component was measured in 
40 of the 42 detections, and the inferred disk masses range
from $0.003-0.07\,M_\odot$.
We find that disks located within 0.3 pc of $\theta^1$\,Ori C have a truncated
mass distribution, while disks located beyond 0.3 pc have masses
more comparable to those found in low-mass star forming regions.
The disk mass distribution in Orion has a distance
dependence, with a derived relationship 
max(M$_{disk}$) = 0.046M$_\odot$(d/0.3pc)$^{0.33}$
for the maximum disk masses.
We found evidence of grain growth in disk 197-427, 
the only disk detected at both $880\,\mu$m and $1330\,\mu$m with the SMA.
Despite the rapid erosion of the outer parts of the Orion disks by photoevaporation,
the potential for planet formation remains high in this massive star 
forming region, with $\approx$\,18\% of the surveyed disks having
masses $\geq\,0.01\,M_\odot$ within 60\,AU.
\end{abstract}


\section{Introduction}\label{sec: intro}
Observations of the circumstellar disks that accompany young 
stars provide important constraints on the planet formation process.  
Their fundamental properties, like mass and size, play a significant role 
in the potential to form planets, and may determine the types of planets that can form.
Our current understanding of disk properties has been expanded through
extensive millimeter wavelength studies of the nearest, young, 
low-mass star forming regions, Taurus-Auriga and $\rho$\,Ophiuchus
\cite[e.g.,][]{beckwith90,osterloh,andre94,andrews05,andrews07}.
The stars in these low-mass star forming regions evolve in relative isolation where only 
internal processes affect their disk properties and evolution, and 
their observations represent an important starting point to understanding
single star and disk evolution.

The vast majority of stars in the galaxy, however, were
not created in regions like Taurus-Auriga or $\rho$\,Ophiuchus, 
but in rich clusters that contain massive ($>8\,M_{\odot}$) O-type stars.
Surveys of low-mass star formation show that 70-90\% of
stars within 2 kpc of the Sun formed in rich, embedded clusters,
$\sim$\,75\% of which currently contain massive stars \citep{lada03}.
Moreover, the mass distribution of molecular clouds is 
weighted toward the most massive clouds, which are more likely to 
form rich clusters containing O-type stars (see \citealt{evans}).
There is even clear evidence that our Sun was born in a massive star
forming environment. Studies of primitive meteorites have
revealed they contain the decay products of $^{60}$Fe, which can only be
produced by a supernova, unambiguously placing
the Sun's formation near at least one massive star \citep{tachibana,krot,gaidos}.

In rich clusters, external influences on disk evolution become
important and they can threaten the development and persistence of 
protoplanetary disks.  The high density of stars can lead to 
enhanced probabilities of dynamical interactions \citep{bonnell03},
while UV radiation from O-type stars can evaporate the disk 
material \citep{johnstone98}. Such is the case in the 
Orion Nebula, which is the nearest, young, massive star forming region.
Hundreds of young stars lie with the central 0.2 pc of the Orion Nebula
Cluster (ONC) \citep{hillenbrand97}, where the
single, $\sim$\,1 Myr old, 45\,$M_{\odot}$, O6-type star, $\theta^1$\,Ori C,
dominates the radiation field of the region.
Normally, protostars like those in Orion would be deeply
embedded leaving them highly obscured throughout their formation.
But the massive OB-stars of the Trapezium cluster, located at the heart of the
Orion Nebula, have carved out a cavity
with their radiation, clearing out the molecular cloud
material along our line of sight, leaving a ``blister HII'' region.
The fortuitous viewing geometry leaves the stars only slightly extincted
($A_v < 2.5$ mag; \citealt{hillenbrand97}), allowing observational access
into the molecular cloud at optical to infrared wavelengths.

The young stars in Orion were initially identified at optical and radio
wavelengths as ionized envelopes surrounding neutral condensations
\citep{laques,moran,garay} and were later hypothesized to be circumstellar 
disks surrounding young stars by \cite{churchwell} through centimeter wavelength 
VLA observations.  \cite{churchwell} measured high mass-loss rates of
$\dot M\sim 10^{-6}-10^{-7}\,M_\odot\,{\rm yr}^{-1}$, which they reasoned 
implied a large reservoir of neutral circumstellar material that would 
render the young stars invisible at optical and infrared wavelengths unless 
it was distributed in a disk-like geometry.  
Hubble Space Telescope (HST) images of the region spectacularly confirmed the disk
hypothesis by imaging the externally illuminated objects and
revealing they were circumstellar disks of dust and gas 
that are evaporating under the intense
UV flux from the most luminous star of the cluster, $\theta^1$\,Ori C 
\citep{odell93,odell94}.

The young objects were named ``proplyds'' by \cite{odell93},
an acryonym for PROtoPLanetarY DiskS, because they appeared
remarkably different from isolated disk systems observed 
in low-mass star forming regions.  
To date, nearly 200 proplyds have been discovered in Orion through 
high resolution optical and near-IR observations with HST 
\citep{odell96,bally98,bally00,smith,ricci}.
Most of the proplyds appear strongly ionized; they are surrounded
by cometary shaped ionized gas cocoons, with bright heads facing
$\theta^1$ Ori C, and tails facing away \citep{mccullough,bally98}.
A small percentage of the disks are dark silhouettes, seen in
extinction against the bright background nebula (see \citealt{mccaughrean96}), or embedded 
silhouettes, projected against the glowing light from their own ionized cocoons. 
The pure dark silhouettes show no evidence of being photoevaporated and are 
likely not in the center of the Trapezium Cluster but only appear so in projection.

Given the mass-loss rates of the disks, 
($\dot M\,\sim\,10^{-6}-10^{-7}\,M_\odot\,{\rm yr}^{-1}$), 
which are high enough to remove a MMSN ($0.01 \,M_\odot$) 
in $\leq\,10^5$\,yr \citep{churchwell,henney}, 
it was questioned whether enough raw material could remain in the
disks to allow planets to form.
However, there is a critical radius surrounding each star, $r_g$,
within which the gravitational potential of the star confines material
to the system for longer than planet formation timescales \citep{johnstone98,adams04}. 
While it takes 0.1 to 1.0 Myrs to remove the outer disk beyond 
$r_g \sim \frac{GM_{\star}}{a^2}$, where $a$ is the sound speed
in the gas, an additional 10-30 Myrs is required to erode the disk
down to 20\,AU scales, into the planet forming zones \citep{adams04,clarke}.
If enough mass exists within $r_g$ of the Orion proplyds, 
then planet formation remains a possibility despite UV photoevaporation.

Although extensive observations have been made of the Orion proplyds at 
optical, infrared and radio wavelengths, none of these observations
were able to yield disk masses.  The disks are purely absorbing at 
optical and near-infrared wavelengths, and therefore, observations were 
only capable of placing lower limits on their column densities and masses, of 
$5 \times 10^{-4}M_\odot = 0.5M_{Jup}$ \citep{mccaughrean98}.
At radio wavelengths, the emission from the proplyds is dominated by 
free-free emission originating in the ionized gas cocoons, which swamps
the weaker dust-disk emission.
In most cases, disk masses can only be estimated from measurements of optically thin
dust continuum emission at millimeter wavelengths.  
The millimeter spectral energy distribution (SED) of a disk
originates from dust located in the cool, disk midplane.
While the infrared observations are sensitive to only $\sim$\,1\% of
the total dust mass (or dust volume), the millimeter observations are sensitive
to dust located out to hundreds of AU, where the majority ($>$90\%) of the mass lies.
Low column densities at these radii make the emission optically thin, 
leading the disk luminosity to be proportional to the sum of emission from
all dust grains. This yields a direct relationship between
submillimeter flux density  F$_{\nu}$ and
disk mass M$_{d}$ \citep{beckwith90}, providing one of the
best ways for masses to be measured.
Disk masses are often compared to a reference called the Minimum
Mass Solar Nebula (MMSN=$0.01 \,M_\odot$),
which represents a lower limit on the amount of gas and dust that was
present in the primordial Solar System disk when the planets began formation
\citep{weidenschilling}.

Since their discovery by HST imaging, the Orion proplyds have been
surveyed by many millimeter wavelength interferometers in an 
attempt to measure their masses and infer their potential to form planets
(BIMA at 3.5-mm, \citealt{mundy}; OVRO at 1.3-mm, \citealt{bally98b};
OVRO at 3-mm, \citealt{eisner06}; CARMA and SMA at 1.3-mm;
\citealt{eisner08}) but few disks were detected due to the limited
sensitivity to dust emission at these long wavelengths.
Even at millimeter wavelengths, the substantial levels of free-free 
emission swamps
the thermal dust emission from the disks, making mass estimates difficult to impossible.
The dust emission increases sharply with frequency, having a dependence
of $F_{\nu} \propto \nu^{2}$, and so the first successful detections of dust 
emission from the Orion proplyds came at {\em sub}-millimeter wavelengths,
$880 \,\mu$m observations with the Submillimeter Array (SMA) \citep{williams}.
Four proplyds were detected by \cite{williams} having disk masses
of 0.013-0.028 M$_\odot$, exceeding the MMSN, and showed the capacity for Orion
disks to be truly protoplanetary.

Building on the success of the pilot study observations by \cite{williams} 
we carried out an SMA survey at $880 \,\mu$m and $1330\,\mu$m
of protoplanetary disks in the Orion Nebula Cluster from late 2006 to early 2010.
The goal of the survey was to measure the disk mass distribution
in a rich cluster in order to the study the impact of massive stars
on disk evolution.
In this paper, we present the results of our complete survey, with observations
taken at $880 \,\mu$m and $1330\,\mu$m, covering 23 SMA fields and 67 proplyds.
We targeted only HST-identified circumstellar disks, as there is no
ambiguity about the geometry of their dust distributions.
Two main results from the survey have been published to date, showing the
inner 0.2 pc of the Orion Nebula lacks the largest
and most massive disks observed in low-mass star forming regions \citep{mann09a},
while the outer regions, beyond 1 pc from $\theta^1$\,Ori C, do not \citep{mann09b}.
Here, we present new SMA observations taken of disks located at intermediate 
distances (d=0.2-1\,pc) from $\theta^1$\,Ori C, to study the disk
mass dependence on distance in the Orion Nebula Cluster.
We describe the observations and data reduction methods in 
$\S$\ref{obs} and present the calculation of disk masses in
$\S$\ref{results}.   
We examine the dependence of disk mass on size and location
within the cluster and also discuss
the results and implications for planet formation in rich clusters
in $\S$\ref{disc}.

\section{Observations}\label{obs}
Submillimeter interferometric observations of 23 fields containing 67 proplyds
were conducted with the Submillimeter Array (SMA; \citealt{ho}) between 2006
Dec 27 and 2010 Jan 29 on Mauna Kea.   
We used the compact and extended array configurations of the SMA;
see Table \ref{table-obs} for a journal of the observations.
The compact configuration was chosen for observations prior to 2009, to provide
the best phase stability, and maintain sufficient resolution
($\sim$ 2.5$^{''}$ at $880\,\mu$m) to distinguish individual proplyds.
Data taken in 2009-2010 were obtained using the extended configuration of the
SMA, which takes advantage of improvements to the array, particularly bandwidth doubling.

\begin{figure}[h]
\centering
\includegraphics[scale=0.9]{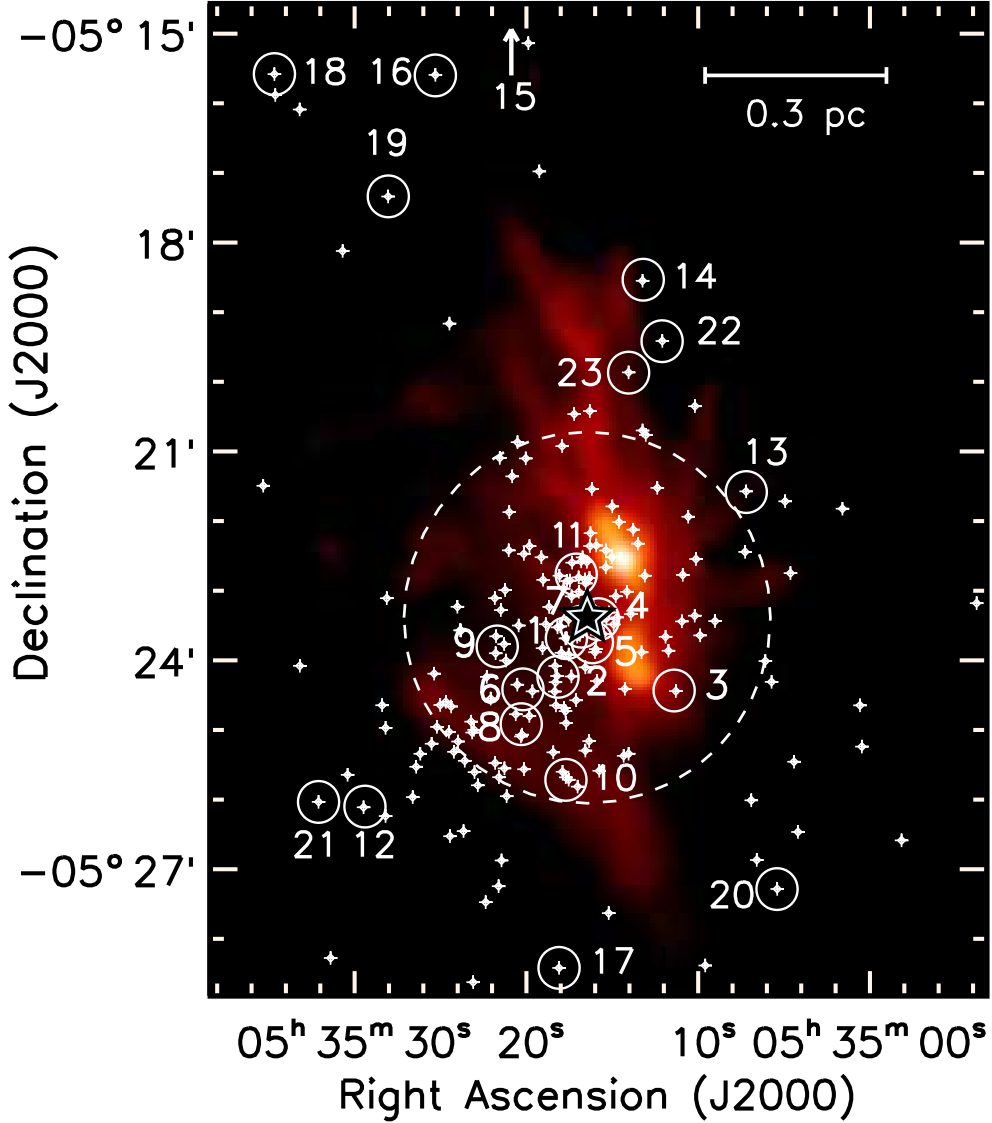}
\caption[Location of the SMA fields overlaid on a SCUBA map]{
Location of the Submillimeter Array (SMA) fields overlaid on a JCMT-SCUBA $850\,\mu$m
image of Orion.  The black and white star near the center
of the image marks the position of $\theta^1$\,Ori C of the Trapezium
Cluster and white crosses show the location of proplyds
identified by HST observations.  The solid white circles represent 
the 32$\arcsec$ primary
beam for SMA observations taken at $880\,\mu$m.
Field numbers are labeled according to Table \ref{table-obs}.  
A dashed circle of radius 0.3 pc, or 155$\arcsec$, around $\theta^1$\,Ori C
represents the sphere of influence of the massive star.
Field 15, which contains proplyd 216-0939, is located a 
few arcminutes North of the image. }
\label{hst+sma}
\end{figure}

The phase centers of the observations (Table \ref{table-obs}, Figure \ref{hst+sma})
were chosen to
simultaneously maximize the number of proplyds imaged while minimizing
contamination from the bright, nonuniform background (see Section \ref{mc}).
Double sideband receivers were tuned to local oscillator (LO) frequencies of
340.175 GHz, 350.175 GHz or 224.170 GHz (882, 857 or 1330 $\mu$m, respectively:
see Table 1).  For observations taken prior to August 2009, each sideband
provided 2 GHz of bandwidth, separated by $\pm$\,5 GHz from the LO frequency.
Subsequent observations benefited from an increased 4 GHz of bandwidth
per sideband, separated by $\pm$\,6 GHz from the LO frequency.
Each observation listed in Table \ref{table-obs} represents a full 8 hour track,
except the last 6 fields (18-23), which were shared 
between pairs of fields because of the increased bandwidth.
For the initial fields, we simultaneously observed the CO(3--2) and HCN(4--3)
transitions by using an LO of 350 GHz.  When it became clear that the line emission from
the disk could not be distinguished from the extended molecular cloud emission,
observations were switched to slightly lower frequencies (340 GHz) where both the
atmosphere and receiver performance improves (see Table \ref{table-obs}). 
CO(3--2), HCN(4--3), CO(2--1) and $^{13}$CO(2--1) were all detected 
but maps show they were indistinguishable from the cloud background 
and we do not discuss them further.

Weather conditions for all observations were good, with $<$ 2 mm
precipitable water vapor, or $\tau$(225 GHz) $<$ 0.1 for $880 \,\mu$m
observations and $<$ 5 mm, or $\tau$(225 GHz) $<$ 0.24 for $1330 \,\mu$m
observations. The atmosphere was very stable during the nights and
system temperatures ranged from 100--400 K.  The total rms noise
levels after calibration ranged from 0.6 to 6.0 mJy.
Table \ref{table-obs} summarizes the relevant observational information.

The raw visibilities for each night were calibrated and edited 
using the MIR software package.
Amplitude and phase calibration were performed through
observations of the bright quasars J0423-013 and J0530+135.
The compact array observations were interleaved between 
20-minute target observations of the proplyds, and 5-minute 
observations of each of the two quasars.
We shortened the target integration to 15-minutes for the 
extended array observations to increase our monitoring of the quasars.
Passband calibration was conducted with one of 
3C454.3, 3C279, or 3C273.  The flux scale was derived using 
measurements of Titan or Uranus as primary flux calibrators;
we used the values derived by Mark Gurwell, available at
http://sma1.sma.hawaii.edu/callist/callist.html.
The flux scale is estimated to be accurate to $\sim$\,15\%.

The calibrated visibilities were weighted by system temperature and
inverted, then cleaned to generate the synthesized continuum maps shown in
Figures \ref{apj1} \& \ref{apj4} using MIRIAD \citep{sault}.
All line transitions were edited out of the observations before producing
the final images.  The maps were created after eliminating $uv$-spacings 
shorter than 27\,k$\lambda$ (physical
baselines shorter than 23-m) to filter out uniform extended emission
corresponding to size scales larger than 7.5$\arcsec$.
The size scale was chosen to preserve compact emission from the disks, 
while minimizing contamination from bright, extended cloud background.
Simulations of the background (see Section \ref{mc}) confirmed
that the 27\,k$\lambda$ cutoff would resolve out the majority of extended
emission and reduce the rms noise levels by factors of 2 to 3.

\section{Results}\label{results}
The 23 SMA fields included a total of 67 HST-identified disks from the
catalogs of \cite{odell96}, \cite{bally00}, \cite{smith}, and \cite{ricci} 
and are shown in Figure \ref{hst+sma}.  
Continuum flux densities, F$_{\rm obs}$, were determined for each 
source by summing the emission within a 2$\sigma$ contour, and 
correcting for primary beam attenuation.  The rms noise levels were 
determined using as much of the emission-free regions within the 
SMA primary beam. In total, we detected 42 of the 
67 surveyed disks with signal-to-noise ratio's $\geq\,3$.

Before disk masses can be calculated, we need to separate
the flux contributions from thermal dust emission from the disks, F$_{\rm dust}$,
the free-free emission from the ionized cocoons, F$_{\rm ff}$, and the background
molecular cloud emission, F$_{\rm bg}$:

\begin{equation}
F_{\rm obs} = F_{\rm dust} + F_{\rm ff} + F_{\rm bg} 
\label{fluxeqn}
\end{equation}

The following sections describe how F$_{\rm ff}$, F$_{\rm bg}$, and
M$_{\rm disk}$ were determined.

\subsection{Free-Free Emission, $F_{\rm ff}$ }
The radio-submillimeter SEDs for the 42 disks 
detected at $\ge$\,3$\sigma$ at $880 \,\mu$m with the SMA 
are shown in Figures \ref{sed1a} and \ref{sed1b}.
We used VLA observations at centimeter wavelengths from
\cite{garay,felli,zapata}, and BIMA, OVRO, CARMA and SMA 
upper limits and detections at millimeter wavelengths
\citep{mundy,bally98b,eisner06,eisner08} to define the
free-free spectrum and then extrapolated it into the 
submillimeter regime. Fits to the free-free emission 
($F_\nu\propto\nu^{-0.1}$) and dust emission
($F_\nu\propto\nu^{2}$) are overlaid on the SEDs to show their 
relative contributions and contrasting spectral dependences.
The long wavelength, 6\,cm to 1.3\,mm, data show a flat spectral 
dependence consistent with optically thin emission, $\nu^{0.1}$, 
but with a range, highlighted by the grey scale, which we 
attribute to variability \citep{felli,zapata}.  
We avoided observations taken at wavelengths longer than 
6\,cm (5 GHz) in this analysis, in order to avoid the turnover
frequency, where the free-free emission becomes 
optically thick and no longer follows a $\nu^{-0.1}$ dependence.
These SEDs show how substantial the free-free
emission towards the Orion proplyds is, swamping the thermal dust emission
from centimeter to millimeter wavelengths, and demonstrating why higher
frequency, submillimeter observations are necessary to detect dust emission
from these disks.  

The disks within fields 12 through 23 
were not detected in previous surveys at radio to millimeter wavelengths and,
with the exception of disk 072-135, none of the disks show signs
of photoevaporation through their HST images, therefore their flux 
contribution from ionized gas emission is expected to be negligible 
at submillimeter wavelengths ($F_{\rm ff} \ll$ 1 mJy ).
F$_{\rm ff}$ is listed in Table \ref{table-det}, and is the maximum
level of free-free radiation extrapolated to the observing frequency.
After subtracting off the free-free contribution to the proplyd fluxes,
40 of the 42 detected proplyds had dust emission in excess of the ionized 
gas emission.  The SMA emission detected towards two proplyds,
168-326NS and 180-331,
is consistent with free-free emission and these are therefore listed
in Table \ref{table-nondet} instead of Table \ref{table-det}.

\subsection{Background molecular cloud emission F$_{\rm bg}$}\label{mc}
The Orion proplyds lie within a blister HII region which sits
in front of a giant molecular cloud.  
The submillimeter emission from the cloud was mapped with the 
SCUBA camera on the James Clerk Maxwell Telescope (JCMT)
by \cite{johnstone99}; see background of Figure \ref{hst+sma}.
The unfiltered background intensity is $\sim$\,1-3 Jy 
per 15$\arcsec$ SCUBA beam, which corresponds to a flux of
$\sim$\,10-30 mJy per compact array SMA beam, or a contribution of 
$\sim\,5-15\,\times\,10^{-3} M_{\odot}$ to a typical disk mass.
The SMA, however, resolves out much of this emission if it is extended.

To characterize the effects of a strong background on the noise 
properties of each field, we performed simulations of the 
interferometric response to the extended cloud emission.
The simulations were conducted using the array configurations 
and $uv$-tracks from the observations and applying them to 
the corresponding positions in the SCUBA map.
For each field, the SCUBA data were Fourier-transformed and
sampled over the observational $uv$-tracks, inverted and cleaned. 
Spatially filtered maps of the background were made using
$uv$-spacings $\geq$ 27 k$\lambda$, to filter out the uniform 
extended emission on angular scales $\geq\,7.5\arcsec$, exactly 
as used to produce the final SMA maps described in
$\S$\ref{obs}.
The flux contribution from the background emission, F$_{\rm bg}$,
was determined by summing the emission 
within an SMA beam sized aperture placed at the position of the individual 
proplyd within the simulated maps.  
The values of F$_{\rm bg}$ are listed in Table \ref{table-det}.
Using the clump mass spectrum in Orion \citep{johnstone01,johnstone06a,johnstone06b},
we estimate the probability of background contamination
from small clumps (7.5$\arcsec$-15$\arcsec$) not
detectable by SCUBA observations to be negligible, with
$\approx$\,10$^{-3}$ detectable clumps per 35$\arcsec$ SMA primary beam.

\subsection{Disk Masses}\label{diskmass}
The submillimeter continuum emission is produced by dust grains
in the disk, which absorb stellar UV and optical photons, and 
re-emit them at longer, infrared to millimeter, wavelengths.
Submillimeter wavelength emission is sensitive to dust
throughout the disk, where the majority ($>$90\%) of the disk mass lies.
Low column densities beyond $\sim$20\,AU imply that the emission is mostly
optically thin and the submillimeter 
fluxes depend directly on the total mass present in the disk:

\begin{equation}
M_{disk} = \frac{d^2F_{\nu}(T)}{\kappa_{\nu}B_{\nu}(T)}.
\label{masseqn}
\end{equation}

The largest uncertainty in this calculation of disk masses 
lies in the value of the dust mass opacity,
$\kappa_{\nu}$, the ratio of effective dust cross section to mass,
which depends strongly on the dust grain characteristics \citep{beckwith90}.

We determined disk fluxes, F$_{\rm dust}$, according to 
Equation \ref{fluxeqn}; see Table \ref{table-det}.
Disk masses were then calculated for the 40 Orion proplyds detected
in significant dust emission using Equation \ref{masseqn}.  We used a distance
to Orion of 400 pc, which is based on recent measurements, including parallax
observations of stars in the cluster, \citep{menten,sandstrom}
and the orbital motion of binaries in the cluster \citep{kraus07,kraus09}.
We used the \cite{beckwith90} dust mass opacity, 
$\kappa_{\nu} = 0.1(\nu/1000$ GHz) cm$^2$g$^{-1}$,
which implicitly assumes a 100:1 gas to dust mass ratio.  
We used a dust temperature of 20 K, the average for disks in Taurus-Auriga and 
$\rho$\,Ophiuchus \citep{andrews05,andrews07}; see also \citet{williams}.
The final disk masses and upper limits for the non-detections 
are listed in Tables \ref{table-det} and \ref{table-nondet}.

The conversion of flux to disk mass is conventional and was chosen for
the simplicity of comparison with other continuum studies, for example,
in Taurus and $\rho$\,Ophiuchus \citep{andrews05,andrews07}.
Longer wavelength observations of Orion disks without substantial free-free emission
show grain growth to beyond millimeter sizes (Ricci et al. 2010b, in press), as is
also seen in Taurus and $\rho$\,Ophiuchus.  The effect lowers the overall
dust opacity, $\kappa_\nu$.
Less direct evidence based on disk accretion rates and ages also suggest that
disk masses may, in general, be underestimated by between a factor of 4-8
\citep{hartmann98,andrews07}.
The disk masses derived here should be considered lower limits but this is 
somewhat mitigated by their comparison to a low estimate of the MMSN.

The mass sensitivity of this survey depends on many factors, including
the varying levels of free-free and background emission contributing
to the observed fluxes, as well as the location of the proplyds within each field.
We, therefore, derived our completeness level by performing
Monte Carlo simulations to determine the fraction of sources that could
be detected at $\geq\,3\sigma$ as a function of mass, depending on the
above-mentioned characteristics.
The results reveal that our survey is 100\% complete for disk masses
M$_d\,\geq\,0.0084\,M_\odot$ and 50\% complete for M$_d\,\geq\,0.003\,M_\odot$
(see also \citealt{mann09a}).

\subsection{$1330\,\mu$m observations}\label{graingrowth}
Although most proplyds appear to be strongly ionized, there are a
small number of silhouette disks which show no evidence of  
photoevaporation through their HST images and have negligible
free-free emission levels at radio wavelengths. 
For these silhouette disks, we attempted to detect dust emission
at a slightly longer wavelength of $1330\,\mu$m.
The detection of dust at more than one wavelength, on
similar baselines, allows us to study the spectral behavior
of the Orion disk emission and constrain their dust properties
by comparing them with disks in Taurus and $\rho$\,Ophiuchus.

In the Rayleigh-Jeans (R-J) limit ($h\nu \ll kT$), the Planck
blackbody function has the dependence $B\,\propto\,\nu^2$,
resulting in the submillimeter flux emission behaving as a
simple power law in frequency, $F_{\nu}\,\propto\,{\nu^{2+\beta}}$.
The power-law index, $\beta$, is a function of the dust grain
size, shape and composition.  If a uniform shape and composition
are assumed, knowledge of $\beta$ can constrain the dust size 
distribution and expose whether the growth of particles has begun in the disk.
The determination of $\beta$ can be complicated if there is significant
deviation from the R-J criterion or if there is a contribution
of optically thick emission to the disk flux \citep{beckwith90}.
For these reasons, we empirically describe the submillimeter continuum emission
as $F_{\nu}\,\propto\,\nu^{\alpha}$, with $\alpha$ related
to $\beta$, but with a minor correction (typically $\Delta\beta\,\sim$\,a few tenths),
for optically thick emission from the inner disk (see \citealt{beckwith90,beckwith91}).

We observed two fields at $1330\,\mu$m, which were centered
on silhouette disks 182-413 and 114-426 (Fields 2 and 3),
neither of which were detected
(see Figure \ref{114} for observations of disk 114-426).
The proplyd 197-427, which lies in the primary beam of Field 2,
was detected at $1330\,\mu$m, with
a primary beam corrected flux of 17.4 mJy
(see Figure \ref{cont230}, Table \ref{grain}).
Disk 197-427 is the only proplyd in Orion detected at 
both $880\,\mu$m and $1330\,\mu$m with our SMA observations.  
The slope of its submillimeter emission, determined from a ratio of its 
$1330\,\mu$m, $880 \,\mu$m fluxes, is  $\alpha$ = 2.8 $\pm$ 0.1.
This slope is intermediate between 
that found for a typical Class II disk in Taurus-Auriga or $\rho$\,Ophiuchus 
with $\alpha\,\sim\,2$ \citep{andrews05,andrews07,rodmann,ricci10}, 
and the ISM, which contains sub-$\micron$ sized dust particles
and has an $\alpha\,\sim$\,4 \citep{pollack94}.
In both Taurus and $\rho$\,Ophiuchus, and recently, in Orion 
(see Ricci et al. 2010b, in press), the systematic change
observed in the submillimeter slope with disk evolutionary state is 
most readily interpreted as due to grain growth in the disks 
to at least millimeter sizes.
Therefore, the slope of Orion disk 197-427 suggests 
dust grain growth may be underway in this disk.
Its spectral slope is actually more consistent with a Class I Taurus
disk than a Class II disk, which we speculate could
be due to photoevaporation of the surrounding circumstellar
envelope, leading to the premature emergence of a Class II disk in Orion.

Disks 182-413, 174-414 and 183-419 were detected at $880 \,\mu$m, 
but not at $1330 \,\mu$m, so we can place constraints on their
spectral slopes using their 3$\sigma$ flux upper limits at $1330 \,\mu$m. 
Fluxes, upper limits and $\alpha$ values for these disks are listed in Table \ref{grain}.
The estimated spectral slopes for Orion disks lie within 
with the range found for disks in Taurus or $\rho$\,Ophiuchus, which show 
$\alpha$=1-4 \citep{andrews05,andrews07,rodmann,ricci10}.

\subsection{Near-Infrared Excess Stars}
Near-infrared emission in excess of stellar photospheric 
emission is often interpreted to be due to the presence
of circumstellar disks around young stars.
\cite{vicente} noted that HST-observations were only able to identify
disks around $\sim$\,50\% of the IR-excess stars in the Orion Nebula Cluster
and suggested that the remaining disks were too small,
$<$\,0.15$\arcsec\approx$\,60\,AU in extent, to be resolved by HST imaging.
In addition to the HST-identified proplyds which were the
primary targets of our survey, there were also 43 near-IR
sources lying within the SMA fields.
With the exception of one source, ID=253 from \cite{hillenbrand00},
which was detected at $\sim$3$\sigma$ in Field 1,
only IR-excess stars with HST-identified counterparts were
detected by the SMA.
The vast majority of the infrared-only disks were not
detected by our observations, implying they are not only 
small in size, but also low in mass.

%---------------------------------------------------------------
\section{Discussion}\label{disc}
\subsection{Non-Detection of the Giant Silhouette Disk 114-426}

Surprisingly, the most prominent Orion disk, 114-426, an 1100\,AU 
silhouette disk seen nearly edge-on in HST images, was not detected 
by the SMA at either $880\,\mu$m or $1330\,\mu$m (see Figure \ref{114}).
The non-detection places a 3$\sigma$ upper mass limit of 
$0.012\,M_\odot$ on the disk at $880\,\mu$m, 
which includes a background flux correction of 
F$_{\rm bg}$ = -12.8 mJy (see Table \ref{table-nondet}).
The previous disk mass upper limit based on a $1330\,\mu$m OVRO non-detection 
\citep{bally98b} is $<\,0.033\,M_\odot$, which we re-calculated using the
improved distance to Orion of 400 pc, and the same T, $\kappa_{\nu}$ as our study.
By studying the extinction of disk 114-426 at multiple wavelengths, 
\cite{mccaughrean98} derived a lower limit on its mass of $5\times10^{-4}\,M_\odot$.
Together, these limits impose a disk mass on 114-426 between
$5\times10^{-4}$-$1.2\times10^{-2}\,M_\odot$ (0.5-12M$_{Jup}$).

The stringent mass upper limit of $10^{-2}\,M_\odot$ on 114-426
is puzzlingly low considering its large extent.
But the non-detection may not be surprising if we consider
the background emission toward this source.
Our SMA simulations, described in Section \ref{mc}, demonstrated that
the ridge of bright background emission that is clearly visible in Figure \ref{hst+sma},
dominates the noise in the
sensitive interferometric observations of the region.
Although we deliberately planned many of our observations
to avoid the bright ridge, we made an exception for 114-426. 
The high contribution from the background emission toward 114-426 is evident in the 
rms noise levels of both $880\,\mu$m and $1330\,\mu$m observations,
which are higher than for pointings not located 
near the bright ridge (see Figure \ref{114}, Table \ref{table-obs}).
The estimated background flux within an SMA beam-sized aperture
toward 114-426 is F$_{\rm bg}$ = -12.8 mJy, the largest absolute
contribution toward any disk in our entire sample (see Tables 
\ref{table-det}, \ref{table-nondet}).
Therefore, the substantial noise levels towards disk 114-426
are not likely due to instrumental issues, but due to small scale
inhomogeneities in the background emission towards this source.
We, therefore, believe that disk 114-426 was not detected by
our SMA observations because of its location near the strong,
contaminating, background molecular cloud emission in the Orion Nebula.

The lack of millimeter emission detected toward disk 114-426 has also 
been attributed to the growth of dust particles to large sizes 
that no longer emit efficiently at these wavelengths \citep{bally98,throop01}.
It is well-established that grain growth is a rapid
process in protoplanetary disks 
\citep[e.g.,][]{dalessio,wilner,dullemond}.
However, grain growth must be
balanced by collisional fragmentation in order to match observations 
that reveal disks remain rich in small dust grains for up to 
several Myrs \citep{dullemond,birnstiel09}. 
It is also difficult to reconcile the grain growth explanation with images 
of the disk taken at $\lambda$=0.4-4$\micron$ \citep{mccaughrean98,throop01,shuping,ricci},
which require a population of $\micron$-sized particles to generate. 
Studies of disk extinction as a function of wavelength
\citep{mccaughrean98,shuping} have even shown that 
most dust grains along the disk edges of 114-426 cannot
be significantly larger than 4-5$\micron$.
Therefore, the primary reason for the non-detection of
disk 114-426 by SMA observations is 
interference from the strong, background emission toward this source,
not grain growth.

\subsection{Disk Masses and Sizes}\label{size}
Theoretical models of disk photoevaporation predict that 
the most extended disks should experience the highest mass-loss
rates because material at larger radii is more loosely bound
to the embedded star and it provides a broader surface area for photoevaporation.
Mass-loss occurs in the disks when the sound speed in 
the gas exceeds the escape velocity:
$ c_s =  (kT/\mu m_H)^{\frac{1}{2}} \geq v_{esc} 
= (GM_{\star}/R)^{\frac{1}{2}}$, or for
disk radii $R_{disk} \ga M_{\star}/T$,
where T is the temperature of the UV-heated disk gas 
and $M_{\star}$ is the stellar mass.
Escape velocities increase towards smaller radii, requiring the disk
gas be heated to higher temperatures, 
$T\,\sim\,1/R_{disk}$, before mass-loss can occur.
Mass-loss rates drop two orders of magnitude, from 
$\dot M\sim10^{-7}\,M_\odot\,{\rm yr}^{-1}$ for 100\,AU disks to
$\sim10^{-9}\,M_\odot\,{\rm yr}^{-1}$ for 20\,AU disks \citep{johnstone98,adams04},
removing the outer disks ($>$\,50-100AU) in 0.1-1.0 Myrs,
while allowing the gravitationally bound, inner ($<$\,50AU) disks
to endure for $\ga$\,10\,Myrs \citep{adams04,clarke}.
The youth of $\theta^1$\,Ori C \citep{hillenbrand97}
implies a photoionizing age $<$\,1\,Myr, and we, therefore,
expect the outer disks of stars located near the Trapezium Cluster
to be stripped away while the inner disks persist.

To investigate the relationship between mass-loss and disk sizes, we
plotted our calculated disk masses versus radii for
31 silhouette disks with resolved sizes
from HST imaging in Figure \ref{diam-mass}; disk radii
were taken from \cite{vicente}.
We did not use the sizes of the unresolved Orion proplyds
from \cite{vicente} in this analysis because they do not have directly 
measured disk radii.
The vertical dashed line in Figure \ref{diam-mass} 
represents the resolution of the HST images,
0.15$\arcsec\approx$ 60\,AU at the distance of Orion.
We expect photoevaporating disks to lose mass and grow
smaller simultaneously, moving them from the upper right
corner of Figure \ref{diam-mass} to the bottom left, 
an effect that is consistent with the observations shown. 

The diversity of initial disk properties within this sample
obscures a clear-cut signature of the relationship
between disk mass and size. 
For example, if we assume all the disks have similar surface
density profiles, with $\Sigma = \Sigma_0 (r/R_0)^{-1}$,
then the disk masses, $M = 2\pi\Sigma_{0}R_0R$, depend not only on
disk size, $R$, but also on the surface density normalization, $\Sigma_0$.
Disk masses and sizes are further complicated by their
dependence on stellar mass, which sets the gravitational 
radius for how far photoevaporation can erode 
the disk on planet formation timescales, and on distance from
$\theta^1$\,Ori C, which also influences the mass-loss rates.
Yet, despite the complexity of the relationship, we find 
evidence of a correlation between disk mass and size.   
The largest disks tend to be the most massive, and there is a lack
of large, low-mass disks.
The probability of correlation between disk mass and radius ranges from
95.4\%-99.95\% ($2.0\sigma - 3.5\sigma$; see Table \ref{asurv}),
which were calculated using various censored statistical tests that
incorporate the 3$\sigma$ upper limits for the non-detections \citep{isobe}.

Spatially resolved observations of many disks in Taurus-Auriga and
$\rho$\,Ophiuchus have revealed that the disk surface density profiles
quite uniformly follow a $\Sigma\,\sim\,r^{-1}$ dependence 
within $\sim$\,100-200\,AU \citep{andrews09,andrews07b}.
We also plot in Figure \ref{diam-mass}, three diagonal dashed lines,
which are comparisons to $M_{disk}$ 
for three normalizations of the surface densities, $\Sigma_0$,
where $\Sigma = \Sigma_0 (r/R_0)^{-1}$.
The uppermost normalization plotted
is the surface density at 5\,AU for a MMSN-disk, 
$\Sigma_{5}$(MMSN$_{60}$), which we define
here to contain $0.01\,M_\odot$ within $R_0$ = 60\,AU:
$\Sigma_{5}$(MMSN$_{60}$) = 
0.01\,M$_\odot$/2$\pi$(60\,AU)(5\,AU) = 47.2 g\,cm$^{-2}$.
The remaining two profiles shown are a factor of $\sqrt10$ and 10 lower.
The photoevaporation of disks should shift them to smaller
sizes and masses along these surface density scalings. 
The normalizations plotted reveal that the majority of Orion silhouette
disks have low surface densities when compared with the standard, MMSN.
We also estimated the normalized disk surface densities at 5\,AU,
$\Sigma_5$, for the Orion silhouette disks to compare them
with Taurus and $\rho$\,Ophiuchus disks (see Figure \ref{density}).
The histograms of log\,($\Sigma_5$), are shown for Orion on top 
and Taurus/$\rho$\,Ophiuchus below, with data taken for the latter
set from \cite{andrews07b}.
The Orion distribution shows a broad peak near a surface density of 
$\sim$\,20\,g\,cm$^{-2}$ at 5\,AU. 
The MMSN surface density at 5\,AU is shown by the dashed line
in Figure \ref{density}.
The distributions between the regions are similar, with many disks
appearing to have lower surface densities than the standard of reference, MMSN.

\subsection{Dependence of Disk Mass on Distance from $\theta^1$\,Ori C}\label{dist}
All theoretical models of protoplanetary disk photoevaporation
by O-stars agree in predicting that the most rapid disk erosion
occurs for the largest disks located near $\theta^1$\,Ori C
\citep{johnstone98,storzer,richling00,scally01,matsuyama,adams04}.
In Orion, the incident radiation field from the massive stars
can be described by its ratio to the interstellar (Habing) field as
$G_0=13,000/d_{pc}^2$ \citep{parravano}.
\cite{adams04} calculated the gas temperatures of photoevaporating
disks under various radiation fields (see their Figure 2)
and showed that the disk temperatures scale roughly with
the radiation field: $T\sim G_0^{\frac{1}{2}}$.  
Therefore, at increasing distances from the massive stars,
UV-heating of the disks declines, $T\sim1/d$,
allowing larger and more massive disks (see previous section)
to survive photoevaporation: $R_{disk} \sim 1/T \sim d$.

In order to examine the dependence of disk mass on location in the cluster,
we plotted our calculated Orion disk masses versus their projected distances
from $\theta^1$\,Ori C in Figure \ref{dist-mass}.
The most massive disks in the cluster are found at the largest
distances from $\theta^1$\,Ori C, while the erosion of the upper
end of the mass distribution towards smaller
distances is clear from this figure. 
It is also interesting to note that all of the non-detections lie
within 0.3 pc from $\theta^1$\,Ori C.
As approximately half of the survey sample were non-detections,
we used censored statistical tests to calculate correlation probabilities.
Results of the statistical tests support a correlation
between Orion disk masses and projected distances from $\theta^1$\,Ori C,
with high probabilities that range between
98.74\% to 100\%; 2.5-4.0\,$\sigma$ (see Table \ref{asurv}).
We used projected distances because 
the true distances of the disks from $\theta^1$\,Ori C
are unknown.  As a consequence, depending on their
radial distance from $\theta^1$\,Ori C, the disks
could in fact lie at larger distances from the central cluster, 
which would shift their positions in Figure \ref{dist-mass} to the right.
However, such an adjustment is not likely to severely alter the trend
of increasing maximum disk masses with distance.
And in spite of the intrinsic diversity of initial disk
properties, which should produce an observable spread in any real correlation,
we find compelling evidence of a dependence between disk mass
and distance from an O-star in the cluster.

The most massive disks are the most readily detectable in our
SMA survey and are subject to the least observational bias.
The maximum disk mass envelope in Orion is traced by the long, dashed line
across the top of Figure \ref{dist-mass}. 
The trend of increasing maximum disk mass with respect 
to distance is obvious, indicative of a declining mass-loss rate.
We derive a power-law relationship to fit the maximum 
mass envelope, which is described by:

\begin{equation}
max(M_{disk}) = 0.046M_\odot\,(\frac{d}{0.3pc})^{0.33}.
\label{maxmass}
\end{equation}

If we assume the disk mass distribution in Orion had no initial
dependence on distance, we can estimate how long it would take 
to photoevaporate it to its present state.
The maximum disk mass at $d\approx$\,1 pc is 0.07\,M$_\odot$, and
at $d\approx$\,0.01 pc is 0.015\,M$_\odot$.
Assuming a mass loss-rate of $\dot M\,\sim\,10^{-7}\,M_\odot\,{\rm yr}^{-1}$
at $d$ = 0.01 pc, it would take $\sim$0.6 Myrs to truncate the
upper end of the disk mass distribution from M$_{disk}$ = 0.07\,M$_\odot$ to 0.015\,M$_\odot$,
which is consistent with the photoevaporation lifetime of $\theta^1$\,Ori C of $<$\,1Myr.

We empirically determined the distance, $d$, at which the MMSN disk
populations within and beyond $d$ differed with the greatest
statistical significance.  
We found 8/14 disks at $d >$\,0.3 pc have masses
$\geq\,0.01\,M_\odot$, while only 11/53 disks within 
0.3 pc have comparable masses. Fisher's exact test
indicates a 96.2\% probability (2$\sigma$) that disks at
large and small projected distances from $\theta^1$\,Ori C have different frequencies.
This distance represents a statistical boundary for circumstellar disk
destruction by external UV photoevaporation %by a nearby O-star 
in the Orion Nebula Cluster.
Orion disks located at $d<$\,0.3 pc have previously been found to have 
a mass distribution that is truncated at its high end (see \citealt{mann09a}).
Beyond 0.3 pc from $\theta^1$\,Ori C, heating by UV radiation is 
insufficient to drive appreciable mass-loss in the disks and we find the
range in disk masses observed, M$_{disk}$ = 0.004-0.07$\,M_\odot$,
is similar to the range exhibited by Class II disks in Taurus and $\rho$\,Ophiuchus.
We, therefore, expect that Orion disks located beyond 0.3 pc from
$\theta^1$\,Ori C evolve similarly to disks in low-mass star
forming regions, Taurus and $\rho$\,Ophiuchus, which do not experience 
UV-driven photoevaporation by nearby massive stars.
Furthermore, it is likely 
that the initial distribution of disk masses in Orion was 
similar to that observed in Taurus-Auriga and $\rho$\,Ophiuchus,
with the Orion disks subsequently photoevaporated from the outside 
inward, and surviving masses and sizes that are strongly dependent 
on their stellar masses and distances from $\theta^1$\,Ori C.

\subsection{Planet Formation Potential}
Despite the clear signs of outer disk destruction observed in Orion, the
prospects for inner disk planet formation are not significantly
different than in low-mass star forming regions.
Overall in Orion, 31\% (21/67) of the disks in our sample have masses
greater than or equal to the MMSN ($0.01\,M_\odot$). For comparison,
the fraction of Class II disks with masses 
$\geq$\,$0.01\,M_\odot$ in Taurus and $\rho$\,Ophiuchus are
$\sim$\,37\% and 29\%, respectively \citep{andrews05,andrews07}.
The fraction of Orion disks with similar properties to the inferred
initial conditions of our Solar System, with masses $\geq\,0.01\,M_\odot$
within 60\,AU, is 12/67 ($\approx$\,18\%), slightly higher than the percentage 
found in Taurus-Auriga of $\approx$\,13\% and comparable to the estimated
fraction of stars with gas giant planets within 20\,AU \citep{cumming}.
Presumably, the lower mass, sub-MMSN disks can form planets less massive than Jupiter,
and including these may raise the prospects for planet formation even further.
While only a third of the total surveyed disks have masses comparable
to a MMSN, all of the disks detected in our survey ($\approx$\,57\%)
have masses $\geq$\,0.28 MMSN.

%---------------------------------------------------------------
\section{Summary}
We have presented the results of our submillimeter interferometric
survey of the 880\,$\micron$ continuum emission from 67 young
circumstellar disks in the Orion Nebula Cluster.  These
observations were taken to study the disk mass distribution 
in the region, and its dependence on distance from the 
most massive star of the cluster, $\theta^1$\,Ori C.  
We find that UV photoevaporation has rapidly ($<$1\,Myr) eroded 
the outer parts of the disks located near this massive star, 
simultaneously reducing their sizes and masses.
We present evidence that shows disk masses correlate with 
distance from $\theta^1$\,Ori C.
Circumstellar disk destruction by UV photoevaporation has a 
statistical limit at the distance of 0.3 pc from $\theta^1$\,Ori C;
beyond 0.3 pc, the flux of UV radiation drops and may not adequately
heat the disk gas to initiate mass-loss.
Despite the hostile environment in the Orion Nebula, the disks
appear very similar to those observed in sites of isolated, low-mass star formation,
Taurus-Auriga and $\rho$\,Ophiuchus, in that we see:

$\bullet$ Similar disk mass distributions after allowances are 
made for photoevaporation of the outer disk edges in Orion.

$\bullet$ Grain growth in one Orion disk, 197-427.

$\bullet$ A similar percentage of disks with 
M\,$\geq$\,0.01$M_{\odot}$ within R\,$\leq$ 60\,AU,
which is also comparable to the estimated 
fraction of radial velocity exoplanets.

We also observed, but did not detect, the largest known 
disk in Orion, 114-426, due to interference from the strong 
background molecular cloud emission toward this source.
The overall potential to form planets in a rich cluster 
containing massive stars like Orion is comparable to that 
found in Taurus-Auriga and $\rho$\,Ophiuchus, 
a promising result in the search for planets like our own.

%---------------------------------------------------------------
\clearpage

\begin{figure}[h]
\centering
\includegraphics[scale=0.7]{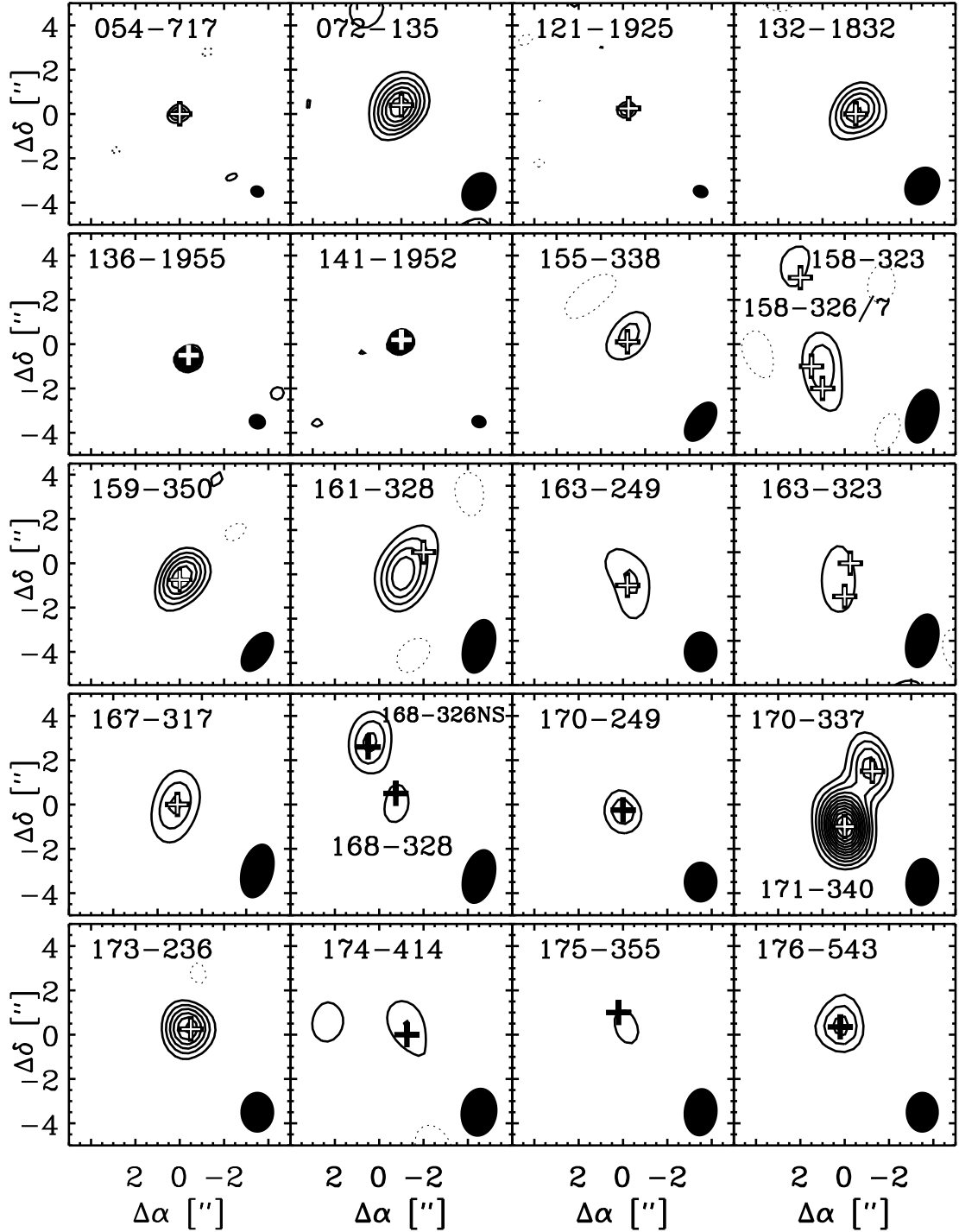}
\vskip 0.1in
\caption[SMA $880\,\mu$m Continuum Observations of Disks in Orion Part I]{
Aperture synthesis images of continuum emission at $880\,\mu$m taken with
the Submillimeter Array towards proplyds the Orion Nebula Cluster.  
All proplyds detected at $\geq$\,3$\sigma$ are shown and labeled
with their names (see also Table \ref{table-det}).
The positions of the HST-identified proplyds are marked by crosses
within the $10\arcsec \times 10\arcsec$ field of view.
Contours begin at the 3$\sigma$ level, and each step represents
2$\sigma$ in intensity (see Table \ref{table-det} for rms noise levels).  
The synthesized beam is shown in the 
bottom right corner of each map.}
\label{apj1}
\end{figure}

\begin{figure}[h]
\centering
\includegraphics[scale=0.7]{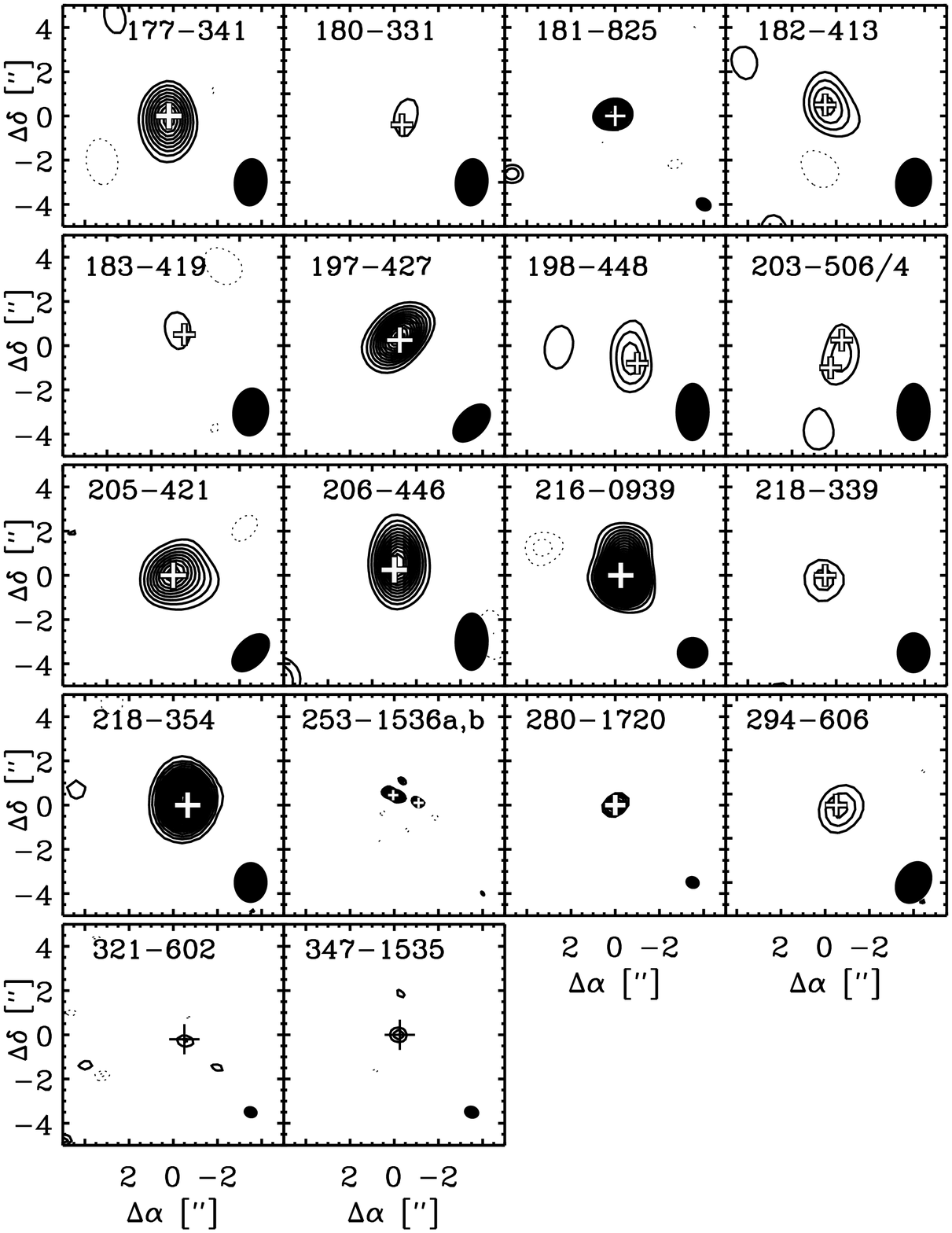}
\vskip 0.15in
\caption[SMA $880\,\mu$m Continuum Observations of Disks in Orion Part II]{
Continued from Figure \ref{apj1}.}
\label{apj4}
\end{figure}

\clearpage
\begin{figure}[h]
\centering
\vskip -0.15in
\includegraphics[scale=0.7]{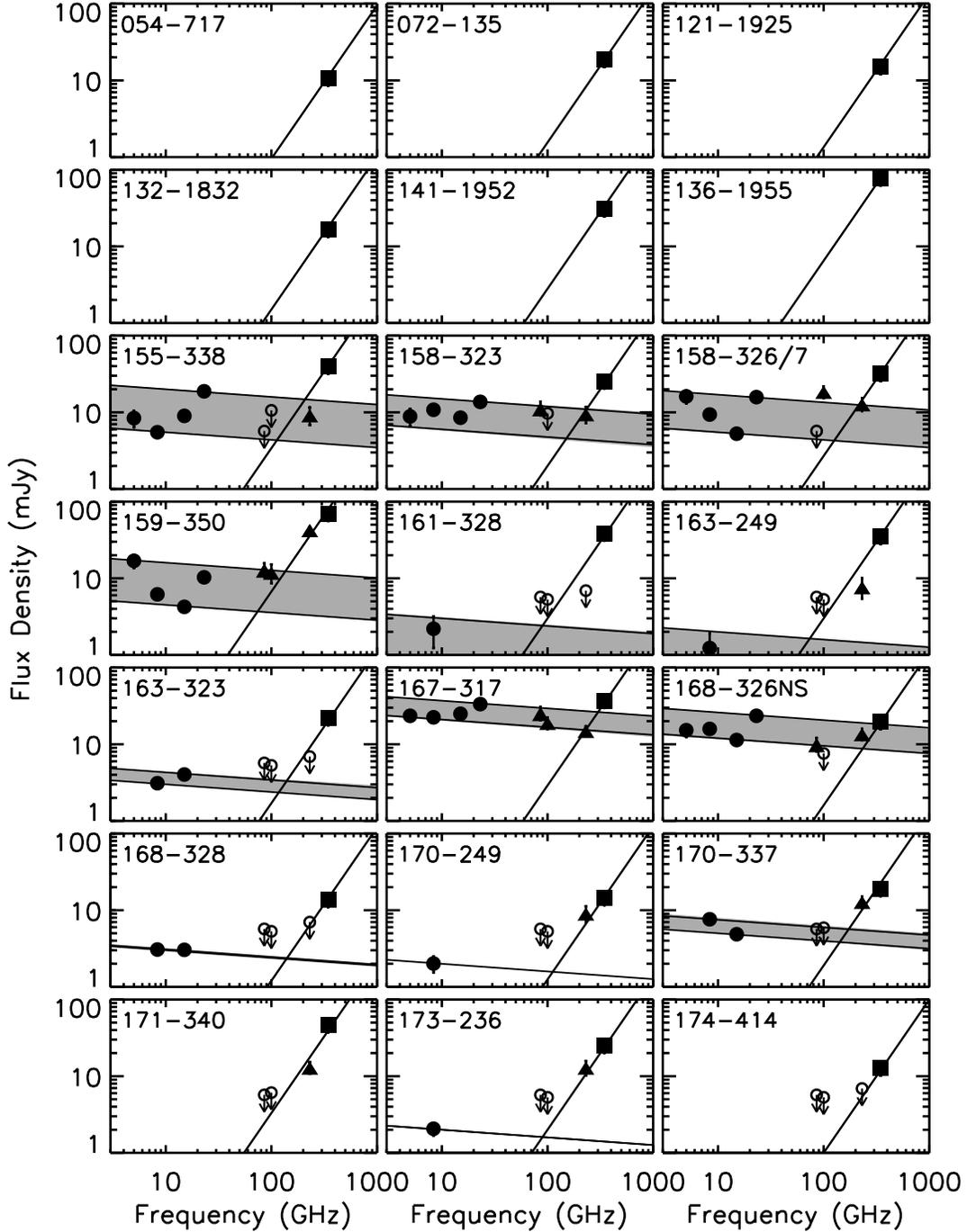}
\caption[Radio-Submillimeter Spectral Energy Distributions of the Proplyds]{
Spectral energy distributions for 42 proplyds detected at
$\geq 3\sigma$ with the Submillimeter Array (SMA) at $880\,\mu$m.
The SMA measurements are represented by squares, millimeter
observations by triangles and centimeter observations by
circles.  Open circles are upper limits from non-detections and
uncertainties not shown are smaller than symbol sizes.
The extrapolated range of optically thin free-free emission,
$F_\nu \propto \nu^{-0.1}$, is overlaid in gray.
A template to the disk emission, $F_\nu \propto \nu^2$, is
shown to guide the eye and reveal the
relative contribution of the ionized gas and dust components.} 
\label{sed1a}
\end{figure} 

\begin{figure}[h]
\centering
\includegraphics[scale=0.75]{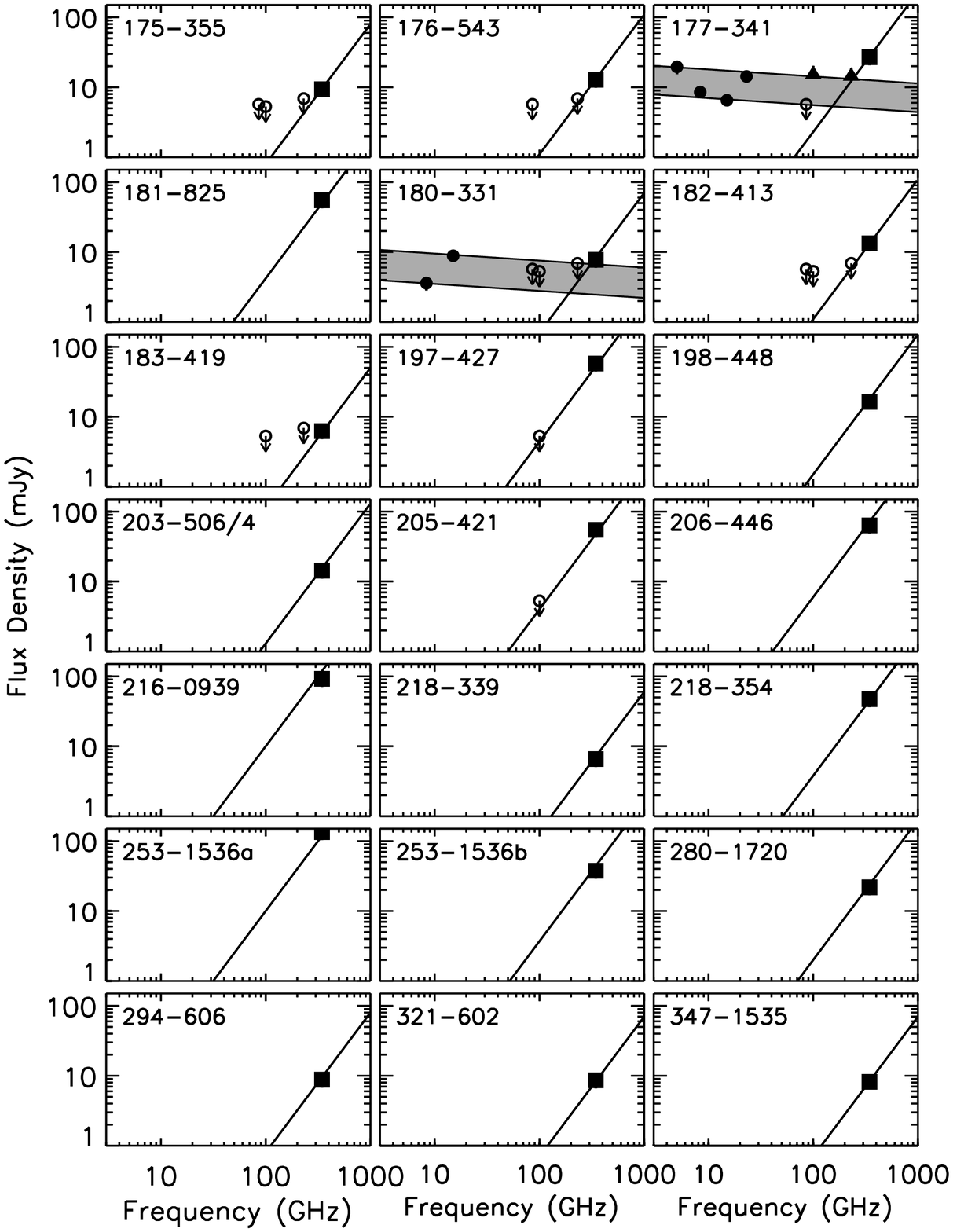}
\caption[Radio-Submillimeter Spectral Energy Distributions of the Proplyds Cont'd.]
{Continued from Figure \ref{sed1a}.}
\label{sed1b}
\end{figure}

\begin{figure}[h]
\centering
\includegraphics[scale=0.82]{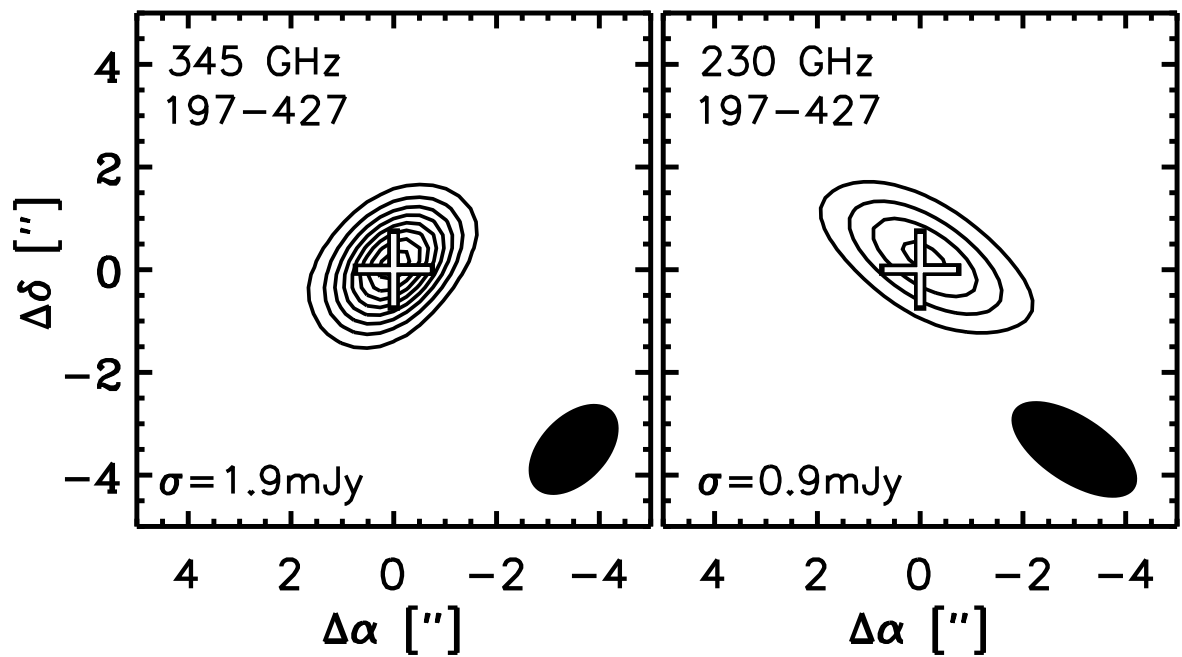}
\caption[SMA $880\,\mu$m and $1330\,\mu$m Continuum Observations of Orion Disk 197-427]{
Aperture synthesis images of continuum emission at 345 GHz ({\em left}) and 
230 GHz ({\em right}) taken with the Submillimeter Array towards prominent 
silhouette disk 197-427 in the Orion Nebula Cluster.  
The rms noise, $\sigma$, is specified in the bottom level corner
of each map.  The contours shown begin at the 3$\sigma$ level and each contour
represents a step of 2$\sigma$ in intensity.  The synthesized
beam is also shown in the bottom right corner of each map. }
\label{cont230}
\end{figure}

\begin{figure}[h]
\centering
\includegraphics[scale=0.46]{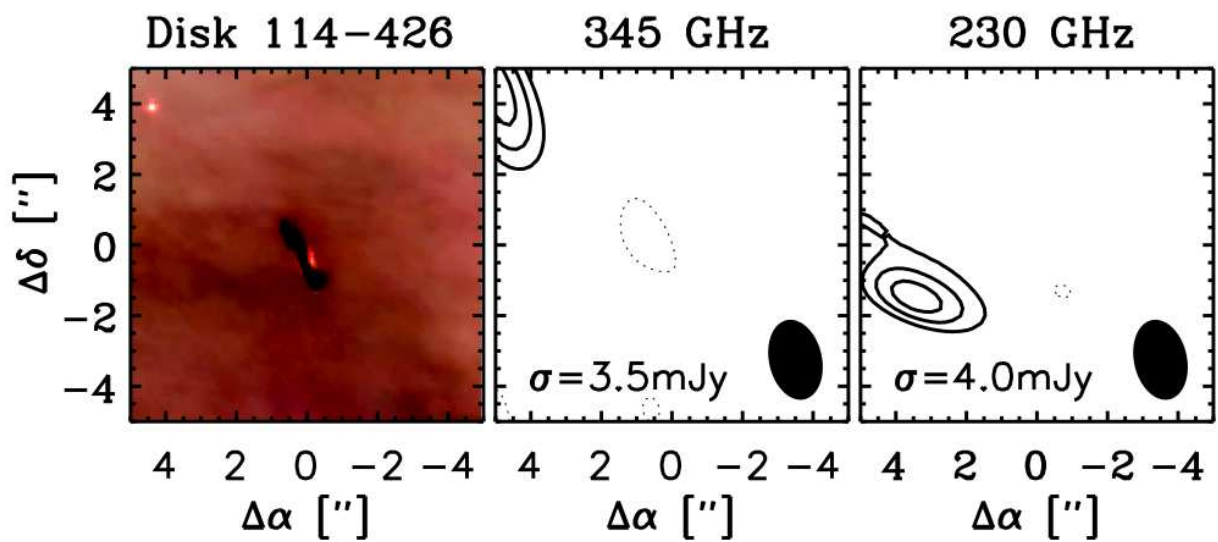}
\vskip -0.75in
\caption[SMA $880\,\mu$m and $1330\,\mu$m Observations of Orion Disk 114-426]{
Hubble Space Telescope (HST) and Submillimeter Array  (SMA)
images of disk 114-426 in Orion.  At left is an HST image
of 114-426 taken from Massimo Robberto.  The middle and right
panels are aperture synthesis images of continuum emission towards
disk 114-426 at 345 GHz ({\em middle}) and 230 GHz ({\em right})
and show the disk was not detected at either wavelength.
The rms noise, $\sigma$, is 
specified in the lower left corner of the SMA maps.  The
contours shown begin at the 2$\sigma$ level and each step represents
1$\sigma$ in intensity.  The synthesized
beam is shown in the bottom right corner of the map.}
\label{114}
\end{figure}

\begin{figure}[h]
\centering
\includegraphics[scale=0.9]{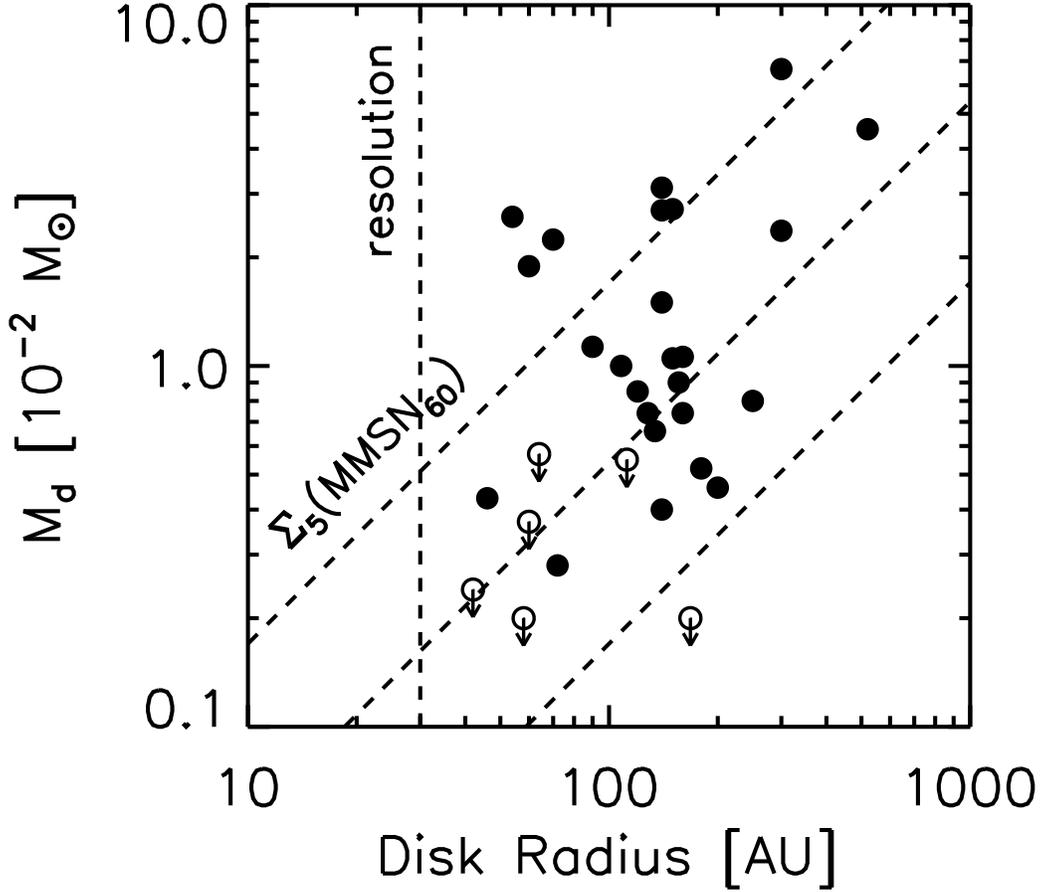}
\caption[Disk Masses and Sizes]{
Disk masses versus radius for the 31 Orion silhouette disks with
resolved sizes from Hubble Space Telescope (HST) imaging.
Filled circles represent detections at $\geq\,3\sigma$,
while open circles are the upper limits for the non-detections.
The dashed vertical line represents the resolution limit of the HST images,
0.15$\arcsec\approx$\,60\,AU.  The diagonal dashed lines represent
different normalizations of the disk surface density, which were calculated
assuming $\Sigma = \Sigma_5$ (r/5\,AU)$^{-1}$.
The top normalization represents a disk mass of $0.01\,M_\odot$ (MMSN)
within 60\,AU, and the remaining profiles are a factor
of $\sqrt10$ and 10 lower. }
\label{diam-mass}
\end{figure}

\begin{figure}[h]
\centering
\vskip -0.15in
\includegraphics[scale=0.8]{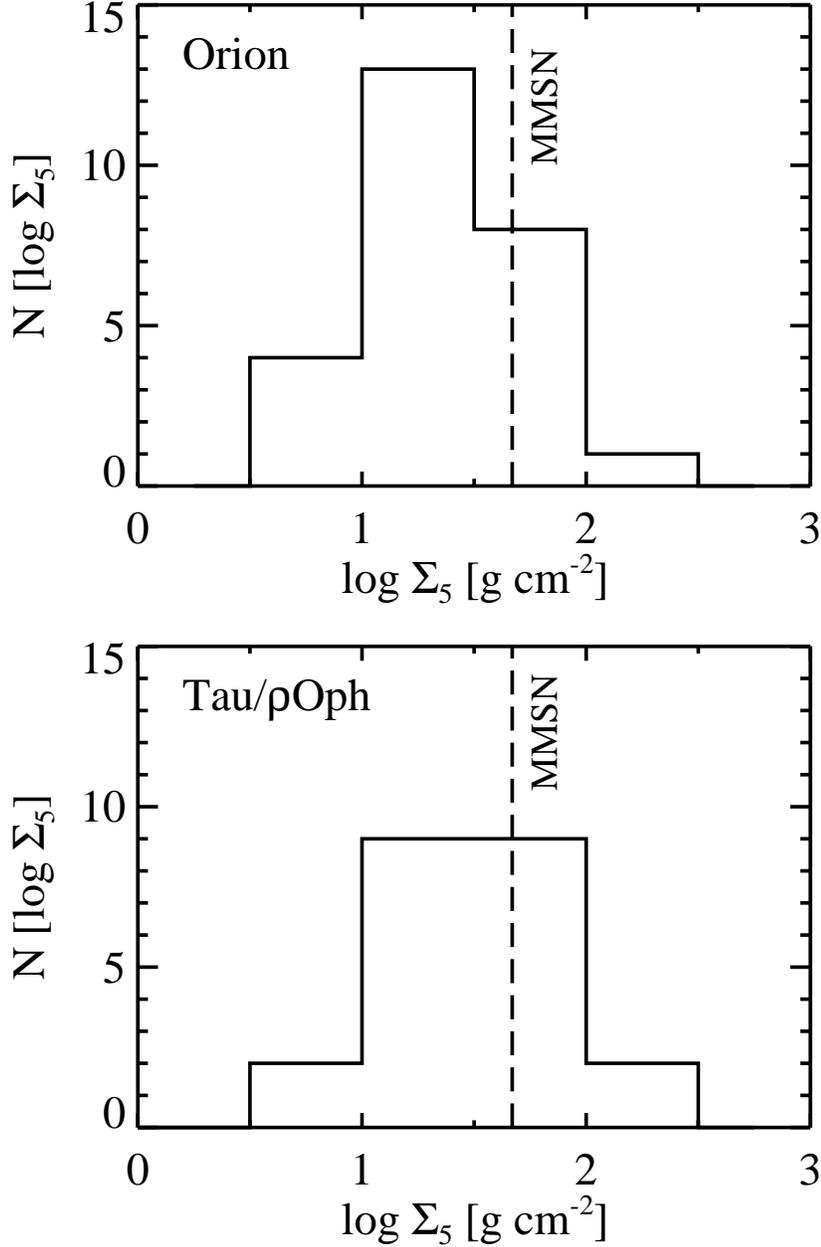}
\caption[Distribution of Normalized Disk Surface Densities]{
The distributions of the normalized disk surface densities at 5\,AU,
log\,$\Sigma_5$ in Orion ({\em top}) and Taurus/$\rho$\,Ophiuchus
({\em bottom}).  For Orion, we show the surface densities for
the sample of 31 silhouette disks with directly
measured sizes from resolved Hubble Space Telescope (HST) images.
For the Taurus/$\rho$\,Ophiuchus distribution, the data were 
taken from \cite{andrews07b}.
The dashed vertical lines represents the
surface density at 5\,AU of the MMSN, 
which was calculated using $\Sigma = \Sigma_5$ (r/5\,AU)$^{-1}$,
and shows the majority of 
disks in both regions have lower surface densities in comparison.}
\label{density}
\end{figure}

\begin{figure}[h]
\centering
\includegraphics[scale=0.84]{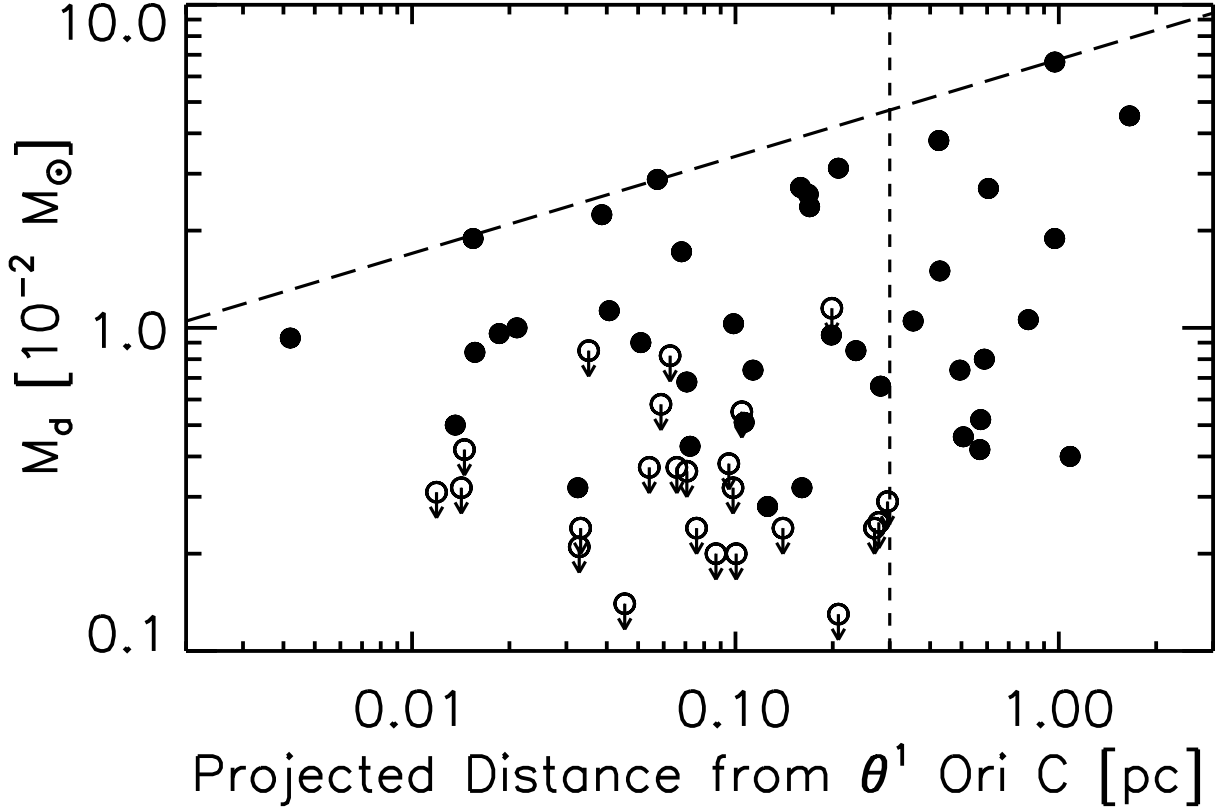}
\caption[Disk Masses as a Function of Distance from $\theta^1$\,Ori C.]{
Circumstellar disk masses plotted against their projected distances
from the massive star, $\theta^1$\,Ori C. Filled circles represent detections,
while open circles are the 3$\sigma$ upper
limits for the non-detections.
The distance of 0.3 pc from $\theta^1$\,Ori C is marked by a dashed line,
to separate the statistically different MMSN-populations.
The maximum disk mass envelope is traced by the long dashed line
across the top, to expose the absence of massive disks near $\theta^1$\,Ori C
and the trend of increasing disk mass with distance.
A derived power-law fit to the maximum mass envelope follows:
max(M$_{disk}$) = 0.046M$_\odot$\,(d/0.3pc)$^{0.33}$.}
\label{dist-mass}
\end{figure}

%---------------------------------------------------------------
% Tables
\clearpage
\begin{deluxetable}{lccccccccr}
\tablecolumns{10}
\tablewidth{0pc}
\tabletypesize{\scriptsize}
\tablecaption{Summary of Submillimeter Array Observations\label{table-obs}}
\tablehead{  \colhead{Field} & \colhead{$\alpha$ (J2000)} &
             \colhead{$\delta$ (J2000)} & \colhead{$\nu_{LO}$ (GHz)} &
             \colhead{UT Date} & \colhead{Array} & 
             \colhead{$\tau$} & \colhead{rms (mJy)} &
             \colhead{Beam ($\arcsec)$} & \colhead{PA($\arcdeg$) }  \\
        \colhead{(1)} & \colhead{(2)} & \colhead{(3)} & \colhead{(4)} &
        \colhead{(5)} & \colhead{(6)} & \colhead{(7)} & \colhead{(8)} &
        \colhead{(9)} & \colhead{(10)} 
}
\startdata
%\hline
1 (a) & 05 35 17.67 & -05 23 40.9 & 350.175 & 2006 Dec 27 & C & 0.03-0.05 & 1.5 & 2.3 x 1.6 & 175.0  \\ % 177-341
2 (c) & 05 35 18.16 & -05 24 14.2 & 350.175 & 2007 Jan 24 & C & 0.04-0.07 & 1.6 & 2.3 x 1.7 & 169.9 \\ % 182-413
2 (c) & 05 35 18.16 & -05 24 14.2 & 224.170 & 2007 Mar 04 & C & 0.17-0.23 & 1.0 & 2.9 x 1.3 & 57.5  \\ % 182-413
3  & 05 35 11.30 & -05 24 26.0 & 224.170 & 2007 Mar 05 & C & 0.23-0.25 & 4.0 & 2.9 x 1.3 & 57.0  \\ % 114-426
3  & 05 35 11.30 & -05 24 26.0 & 340.175 & 2008 Oct 22 & C & 0.08-0.09 & 3.5 & 2.3 x 1.5 &  12.6 \\ % 114-426
4  & 05 35 15.80 & -05 23 24.5 & 349.385 & 2007 Nov 17 & C & 0.08-0.1  & 4.5 & 1.9 x 1.4 & 137.4 \\ % PROPLYD A
5 (e) & 05 35 20.20 & -05 24 25.0 & 349.379 & 2007 Nov 25 & C & 0.06-0.1  & 1.9 & 2.2 x 1.4 & 135.3 \\ % 197-427
6 (d) & 05 35 16.16 & -05 23 44.5 & 340.175 & 2007 Dec 18 & C & 0.05-0.06 & 5.8 & 2.3 x 1.5 & 145.3 \\ % Prop B
7 (b) & 05 35 16.80 & -05 23 31.0 & 340.175 & 2008 Jan 08 & C & 0.03-0.04 & 2.2 & 2.7 x 1.6 & 164.2 \\ % Prop C
8 (f) & 05 35 20.30 & -05 24 55.0 & 340.175 & 2008 Mar 14 & C & 0.09-0.11 & 2.0 & 2.6 x 1.5 & 177.8 \\ % 206-446
9 (h) & 05 35 21.70 & -05 23 48.0 & 340.175 & 2008 Sep 17 & C & 0.05-0.06 & 0.7 & 1.8 x 1.5 &   1.6 \\ % 218-354
10 (g) & 05 35 17.70 & -05 25 43.0 & 340.175 & 2008 Sep 24 & C & 0.05-0.15 & 1.4 & 2.0 x 1.6 &   3.0 \\ % 176-543
11 (i) & 05 35 17.10 & -05 22 46.0 & 340.175 & 2008 Sep 30 & C & 0.05-0.15 & 2.0 & 1.8 x 1.5 & 176.4 \\ % 170-249
12    & 05 35 29.40 & -05 26 06.0 & 340.175 & 2008 Nov 11 & C  & 0.06-0.08 & 1.2 & 2.2 x 1.8 & 150.4 \\ % 294-606
13    & 05 35 07.20 & -05 21 35.0 & 340.175 & 2008 Nov 26 & C  & 0.08-1.0  & 1.7 & 2.1 x 1.6 & 148.0 \\ % 072-135
14    & 05 35 13.20 & -05 18 32.0 & 340.175 & 2008 Dec 21 & C  & 0.06-0.07 & 1.7 & 1.9 x 1.5 & 139.1 \\ % 132-1832
15    & 05 35 21.60 & -05 09 39.0 & 340.175 & 2008 Dec 23 & C  & 0.10-0.15 & 1.3 & 2.1 x 1.6 & 146.1 \\ % 1.9 x 1.4; 51.4
16    & 05 35 25.30 & -05 15 36.0 & 340.175 & 2008 Dec 24 & C  & 0.13-0.16 & 2.3 & 1.9 x 1.4 & 157.3 \\ % 253-1536
16    & 05 35 25.30 & -05 15 36.0 & 340.175 & 2009 Mar 26 & V  & 0.05-0.10 & 1.0 & 0.3 x 0.2 & 35.0 \\ % VEX
16    & 05 35 25.30 & -05 15 36.0 & 340.175 & 2009 Mar 27 & V  & 0.03-0.05 & 1.0 & 0.3 x 0.2 & 35.0 \\ % VEX
17    & 05 35 18.10 & -05 28 25.0 & 340.175 & 2009 Aug 25 & E  & 0.04-0.06 & 1.0 & 0.9 x 0.7 & 110.1 \\ % Beehive
18    & 05 35 34.67  & -05 15 34.7 & 340.765 & 2010 Jan 19 & E & 0.02-0.03 & 1.4 & 0.7 x 0.6 & 110.0  \\
19    & 05 35 28.04  & -05 17 20.0 & 340.765 & 2010 Jan 19 & E & 0.02-0.03 & 0.7 & 0.8 x 0.6 & 122.6  \\
20    & 05 35 05.40  & -05 27 17.1 & 342.000 & 2010 Feb 28 & E & 0.06-0.08 & 1.4 & 0.8 x 0.7 & 106.6  \\
21    & 05 35 32.10  & -05 26 02.0 & 342.000 & 2010 Feb 28 & E & 0.06-0.08 & 1.5 & 0.7 x 0.5 &  97.8  \\
22    & 05 35 12.10  & -05 19 25.0 & 340.765 & 2010 Feb 29 & E & 0.05-0.08 & 1.5 & 0.7 x 0.6 & 109.9  \\
23    & 05 35 14.05  & -05 19 52.1 & 340.765 & 2010 Feb 29 & E & 0.05-0.08 & 1.2 & 0.8 x 0.7 &  69.7  \\
\enddata
\tablecomments{
{\sc Notes} --- Column 1: Field Number; also labeled in Figure \ref{hst+sma}.  
Previous designation from \citep{mann09b} listed in brackets.
Column 2, 3: Phase Center Coordinates. Column 4: Observing Frequency.  
Column 5: UT Date of Observation.  Column 6: Array configuration; C, compact (16-70 m baselines), 
E, extended (28-226 m baselines), V, very extended (68-509 m baselines).
Column 7: Zenith optical depth at 225 GHz. 
Column 8: Root mean square noise measured in emission-free regions within the primary beam.
Column 9: Dimensions of the naturally weighted synthesized beam.
Column 10: Position angle of synthesized beam, measured east of north.}
\end{deluxetable}

\clearpage
\vskip -1in
\begin{deluxetable}{lccccccr}
\tablecolumns{8}
\tablewidth{0pc}
\tabletypesize{\scriptsize}
\tablecaption{Disk Fluxes and Masses\label{table-det}}
\tablehead{  \colhead{Proplyd} & \colhead{Field} &
             \colhead{F$_{\rm obs}$} & \colhead{rms} &
             \colhead{F$_{\rm ff}$} & \colhead{F$_{\rm bg}$} &
             \colhead{F$_{\rm dust}$} & \colhead{M$_{\rm disk}$} \\

        \colhead{} & \colhead{} & \colhead{[mJy]} & \colhead{[mJy]} &
        \colhead{[mJy]} & \colhead{[mJy]} & \colhead{[mJy]} & \colhead{[0.01M$_{\odot}$]} \\

        \colhead{(1)} & \colhead{(2)} & \colhead{(3)} & \colhead{(4)} &
        \colhead{(5)} & \colhead{(6)} & \colhead{(7)} & \colhead{(8)} 
}
\startdata
 054-717   & 20 & 10.7  & 1.4 &    0.0    &  0.0 &  10.7  & 0.52 $\pm$ 0.07  \\
 072-135   & 13 & 18.6  & 1.7 &    0.0    & -2.8 &  11.5  & 1.05 $\pm$ 0.08 \\
 121-1925  & 22 & 15.0  & 1.5 &    0.0    &  0.0 &  15.0  & 0.74 $\pm$ 0.07  \\
 132-1832  & 14 & 16.5  & 1.7 &    0.0    &  0.2 &  16.3  & 0.80 $\pm$ 0.08 \\
 136-1955  & 23 & 77.6  & 1.2 &    0.0    &  0.0 &  77.6  & 3.80 $\pm$ 0.07  \\
 141-1952  & 23 & 30.6  & 1.2 &    0.0    &  0.0 &  30.6  & 1.50 $\pm$ 0.06  \\
155-338    & 5  & 39.6  & 5.3 & 4.0--14.0 &  2.6 &  23.0  & 1.13 $\pm$ 0.34  \\ %B
158-323    & 7  & 25.4  & 2.4 & 4.3--11.5 & -5.6 &  19.5  & 0.96 $\pm$ 0.21 \\  %C
158-326/7  & 7  & 31.9  & 2.4 & 4.0--12.0 & -0.5 &  20.3  & 1.00 $\pm$ 0.19 \\
159-350    & 5  & 69.8  & 5.3 & 3.2--11.0 &  0.3 &  58.4  & 2.88 $\pm$ 0.29 \\
161-328    & 7  & 38.7  & 2.4 & 0.1--1.5  & -1.2 &  38.4  & 1.89 $\pm$ 0.15 \\
163-249    & 11 & 34.8  & 2.0 &    0.0    & -1.6 &  35.0  & 1.72 $\pm$ 0.13 \\
163-323    & 7  & 22.1  & 2.4 & 2.4--3.0  &  0.2 &  18.9  & 0.93 $\pm$ 0.15 \\
167-317    & 7  & 37.2  & 2.4 & 14.5--25.5& -5.4 &  17.0  & 0.84 $\pm$ 0.18 \\
168-328    & 7  & 13.7  & 2.4 & 1.9--2.5  &  1.0 &  10.3  & 0.50 $\pm$ 0.12 \\
170-249    & 11 & 14.6  & 2.0 & 0.0--2.5  & -0.6 &  13.7  & 0.68 $\pm$ 0.10 \\
170-337    & 1  & 19.1  & 1.5 & 4.0--9.0  &  3.0 &   7.1  & 0.32 $\pm$ 0.08 \\
171-340    & 1  & 46.4  & 1.5 &    0.0    & -2.8 &  49.2  & 2.24 $\pm$ 0.08 \\
173-236    & 1  & 25.2  & 2.0 &    0.0    &  2.9 &  20.8  & 1.03 $\pm$ 0.13 \\
174-414    & 2  & 12.7  & 1.5 &    0.0    &  1.6 &  11.1  & 0.51 $\pm$ 0.09 \\
175-355    & 8  &  9.4  & 1.5 &    0.0    & -0.1 &   9.5  & 0.43 $\pm$ 0.11 \\
176-543    & 10 & 12.8  & 1.4 &    0.0    & -0.6 &  13.4  & 0.66 $\pm$ 0.07 \\ % F
177-341    & 1  & 26.9  & 1.5 & 5.0--13.0 & -5.9 &  19.8  & 0.90 $\pm$ 0.07 \\
 181-825   & 17 & 54.8  & 1.0 &    0.0    &  0.0 &  54.8  & 2.70 $\pm$ 0.05 \\
182-413    & 2  & 13.3  & 1.5 &    0.0    & -3.0 &  16.4  & 0.74 $\pm$ 0.07 \\
183-419    & 2  &  6.2  & 1.5 &    0.0    &  0.2 &   6.1  & 0.28 $\pm$ 0.07 \\
197-427    & 6  & 57.7  & 1.9 &    0.0    & -1.5 &  59.3  & 2.72 $\pm$ 0.10 \\
198-448    & 8  & 16.4  & 2.0 &    0.0    & -2.9 &  19.3  & 0.95 $\pm$ 0.12 \\
203-506/4  & 8  & 14.3  & 2.0 &    0.0    & -2.9 &  17.2  & 0.85 $\pm$ 0.13 \\
205-421    & 6  & 54.8  & 1.9 &    0.0    & -1.7 &  56.5  & 2.59 $\pm$ 0.10 \\
206-446    & 8  & 63.2  & 2.0 &    0.0    &  0.0 &  63.3  & 3.12 $\pm$ 0.12 \\
 216-0939  & 15 & 91.9  & 1.3 &    0.0    &  0.0 &  91.9  & 4.53 $\pm$ 0.06 \\
218-339    & 9  &  6.6  & 0.7 &    0.0    &  0.0 &   6.6  & 0.32 $\pm$ 0.04 \\
218-354    & 9  & 42.3  & 0.7 &    0.0    & -0.9 &  48.1  & 2.37 $\pm$ 0.04 \\ %G
 253-1536a & 16 &134.2  & 1.0 &    0.0    & -0.8 & 135.0  & 6.65 $\pm$ 0.05 \\
 253-1536b & 16 & 37.6  & 1.0 &    0.0    & -0.8 &  38.4  & 1.89 $\pm$ 0.05 \\
 280-1720  & 19 & 21.8  & 0.7 &    0.0    &  0.0 &  21.8  & 1.06 $\pm$ 0.03  \\
 294-606   & 12 &  8.8  & 1.2 &    0.0    & -0.5 &   9.3  & 0.46 $\pm$ 0.06 \\
 321-602   & 21 &  8.6  & 1.5 &    0.0    &  0.0 &   8.6  & 0.42 $\pm$ 0.07  \\
 347-1535  & 18 &  8.2  & 1.4 &    0.0    &  0.0 &   8.2  & 0.40 $\pm$ 0.07  \\
\enddata
\tablecomments{
{\sc Notes} --- Column 1: Proplyd designation based on the nomenclature of O'Dell \& Wen (1994).
Column 2: Observed Field in Figure \ref{hst+sma} and Table \ref{table-obs}.
Column 3: Integrated continuum flux density, corrected for SMA primary beam attenuation.
Column 4: 1$\sigma$ statistical error.
Column 5: Range of extrapolated contribution of free-free emission at $880 \,\mu$m.
Column 6: Estimated flux contribution from cloud background.
Column 7: Derived dust continuum flux from the disk.
Column 8: Disk mass (error does not include uncertainties in the flux scale of $\sim$ 15\%). }
\end{deluxetable}

\clearpage
\begin{deluxetable}{lccccr}
\tablecolumns{6}
\tablewidth{0pc}
\tabletypesize{\scriptsize}
\tablecaption{Upper Limits on Disk Fluxes and Masses\label{table-nondet}}
\tablehead{  \colhead{Proplyd} & \colhead{Field} &
             \colhead{3\,$\sigma$} &  \colhead{F$_{\rm ff}$} & \colhead{F$_{\rm bg}$} &
             \colhead{M$_{\rm disk}$} \\

        \colhead{} & \colhead{} & \colhead{[mJy]} & \colhead{[mJy]} &
        \colhead{[mJy]} & \colhead{[0.01M$_{\odot}$]} \\

        \colhead{(1)} & \colhead{(2)} & \colhead{(3)} & \colhead{(4)} &
        \colhead{(5)} & \colhead{(6)} 
}
\startdata
114-426   & 3  & $<$ 10.5  &  0        & -12.8  &  $<$ 1.15      \\
159-338   & 6  & $<$ 17.4  & 0.0--1.2  & -0.9   &  $<$ 0.85      \\
160-353   & 6  & $<$ 19.3  & 0.0--2.0  &  0.5   &  $<$ 0.82      \\
161-324   & 7  & $<$ 10.0  & 2.5--4.5  & -0.8   &  $<$ 0.31      \\
163-317   & 7  & $<$ 12.8  & 4.6--12.0 & -5.7   &  $<$ 0.32      \\
165-235   & 11 & $<$  9.3  &  0        &  1.6   &  $<$ 0.38      \\
165-254   & 11 & $<$  8.0  &  0        & -3.7   &  $<$ 0.57      \\
166-250   & 11 & $<$  7.1  &  0        & -0.3   &  $<$ 0.37      \\
166-316   & 7  & $<$ 11.6  & 0.1--1.5  &  1.7   &  $<$ 0.42      \\
167-231   & 11 & $<$ 10.5  &  0        & -0.6   &  $<$ 0.55      \\
168-235   & 11 & $<$  8.4  &  0        &  2.0   &  $<$ 0.32      \\
168-326NS & 7  &   19.6    & 8.6--18.0 &  1.2   &  $<$ 0.02      \\
169-338   & 1  & $<$  6.0  &  0        &  0.7   &  $<$ 0.24      \\
169-549   & 10 & $<$  4.6  &  0        & -1.3   &  $<$ 0.29      \\
171-334   & 1  & $<$  5.7  & 1.9--4.7  &  1.2   &  $<$ 0.07      \\
173-341   & 1  & $<$  4.6  &  0        &  1.6   &  $<$ 0.14      \\
175-251   & 11 & $<$  6.7  &  0        & -0.5   &  $<$ 0.36      \\
176-325   & 7  & $<$ 10.2  & 2.8--4.0  &  1.9   &  $<$ 0.21      \\
177-541   & 10 & $<$  4.3  &  0        & -0.8   &  $<$ 0.25      \\
179-353   & 1  & $<$  6.5  &  0        &  1.3   &  $<$ 0.24      \\
179-534   & 10 & $<$  4.7  &  0        & -0.2   &  $<$ 0.24      \\
180-331   & 1  &  7.7      & 2.1--6.7  &  2.1   &  $<$ 0.07      \\
181-247   & 11 & $<$  9.3  & 1.6--3.2  &  0.9   &  $<$ 0.20      \\
182-332   & 1  & $<$  6.1  &  0        & -2.1   &  $<$ 0.37      \\
183-405   & 2  & $<$  5.9  &  0        &  1.5   &  $<$ 0.20      \\
184-427   & 2  & $<$  7.0  &  0        &  1.7   &  $<$ 0.24      \\
213-346   & 9  & $<$  2.6  &  0        &  0.0   &  $<$ 0.13      \\
\enddata
\tablecomments{
{\sc Notes} --- Column 1: Proplyd designation based on the nomenclature of O'Dell \& Wen (1994).
Column 2: Observed Field in Figure \ref{hst+sma} and Table \ref{table-obs}.
Column 3: 3\,$\sigma$ flux upper limits (or integrated continuum flux density for
168-326N, 180-331), corrected for primary beam attenuation.
Column 4: Extrapolated contribution of free-free emission at 880$\mu$m.
Column 5: Estimated flux contribution from cloud background at 880$\mu$m.
Column 6: 3\,$\sigma$ disk mass upper limit. }
\end{deluxetable}

\begin{deluxetable}{lcccr}
\tablecolumns{5}
\tablewidth{0pc}
\tabletypesize{\scriptsize}
\tablecaption{Submillimeter Continuum Slopes\label{grain}}
\tablehead{  \colhead{Proplyd} & \colhead{Field} &
             \colhead{F$_{\rm 880 \mu m}$} &  \colhead{F$_{\rm 1330\mu m}$} & \colhead{$\alpha$} \\
        \colhead{} & \colhead{} & \colhead{[mJy]} & \colhead{[mJy]} & \colhead{} \\
        \colhead{(1)} & \colhead{(2)} & \colhead{(3)} & \colhead{(4)} & \colhead{(5)} 
}
\startdata
197-427 & 2 & 57.7 & 17.4 & 2.8 $\pm$ 0.1 \\
174-414 & 2 & 12.7 & $<$3.4 & $>$ 3.0 $\pm$ 0.1 \\
182-413 & 2 & 13.3 & $<$3.0 & $>$ 3.4 $\pm$ 0.1 \\
183-419 & 2 &  6.2 & $<$3.1 & $>$ 1.6 $\pm$ 0.1 \\
114-426 & 3 & $<$10.5 & $<$12.0 &   \\
\enddata
\tablecomments{
{\sc Notes} --- Column 1: Proplyd designation based on the nomenclature of O'Dell \& Wen (1994).
Column 2: Observed Field in Figure \ref{hst+sma}.
Column 3: Integrated continuum flux density at 880\,$\mu$m, corrected for primary beam attenuation.
Column 4: Integrated continuum flux density at 1330\,$\mu$m, corrected for primary beam attenuation.
Column 5: Submillimeter SED Slope as defined in Section \ref{graingrowth}.
%$^*$ Fluxes were measured using baselines longer than 70 k$\lambda$ for this disk, to allow 
comparison of SMA and CARMA observations. }
\end{deluxetable}

\begin{deluxetable}{lcr}
\tablecolumns{3}
\tablewidth{0pc}
\tabletypesize{\scriptsize}
\tablecaption{Statistical Correlation Tests of Orion Disk Properties\label{asurv}}
\tablehead{  \colhead{Correlaton Test} & \colhead{Mass vs. Size} & \colhead{Mass vs. Distance} \\
        \colhead{} & \colhead{P(\%)} & \colhead{P(\%)} \\
        \colhead{(1)} & \colhead{(2)} & \colhead{(3)} 
}
\startdata
Cox Hazard              &       99.95 (3.5$\sigma$) &  99.99 (4.0$\sigma$) \\
Kendall's $\tau$        &       95.36 (2.0$\sigma$) &  98.74 (2.5$\sigma$)\\
Spearman's $\rho$       &       97.17 (2.2$\sigma$) &  99.07 (2.6$\sigma$) \\
\enddata
\tablecomments{
{\sc Notes} --- Column 1: 
Censored statistical correlation tests were used that incorporate 3$\sigma$ upper limits for the
non-detections \citep{isobe}. These tests include the Cox Proportional Hazard Model,
the Generalized Kendall's $\tau$, and Spearman's $\rho$.
Column 2: Probability of correlation between disk size and mass for silhouette disks.
Column 3: Probability of correlation between disk mass and the projected distance
of the disk from $\theta^1$\,Ori C. }
\end{deluxetable}

%---------------------------------------------------------------
\clearpage
{\bibliographystyle{apj}}
{\bibliography{apj-jour,bib_rm}}

\end{document}